\documentclass[pra,twocolumn,aps,showpacs,amsmath,amssymb,superscriptaddress,floatfix]{revtex4-2} 

\usepackage{dcolumn}
\usepackage{braket}
\usepackage{lipsum}
\usepackage{multirow}

\usepackage{graphicx}
\usepackage{mathtools}
\usepackage[export]{adjustbox}
\usepackage{epsfig}
\usepackage{color}
\usepackage{subfigure}
\usepackage{setspace}
\usepackage{bm}
\usepackage{calc}
\usepackage{natbib}
\usepackage[normalem]{ulem}
\usepackage{cancel}
\usepackage{soul}

\usepackage[pdftex]{hyperref}
\usepackage{ulem} 
\hypersetup{colorlinks = true, urlcolor = blue, linkcolor = blue, citecolor = blue}

\usepackage{dblfloatfix}
\usepackage{flushend} 

\usepackage{xcolor}
\usepackage{xparse,xcoffins}
\ExplSyntaxOn
\NewCoffin\imagecoffin
\NewCoffin\labelcoffin
\keys_define:nn { miguel/label }
 {
  label   .tl_set:N = \l_miguel_label_tl,
  labelbox .bool_set:N = \l_miguel_label_box_bool,
  labelbox .default:n = true,
  fontsize .tl_set:N = \l_miguel_label_size_tl,
  fontsize .initial:n = \footnotesize,
  pos .choice:,
  pos/nw .code:n = \tl_set:Nn \l_miguel_label_pos_tl { left,up },
  pos/ne .code:n = \tl_set:Nn \l_miguel_label_pos_tl { right,up },
  pos/sw .code:n = \tl_set:Nn \l_miguel_label_pos_tl { left,down },
  pos/se .code:n = \tl_set:Nn \l_miguel_label_pos_tl { right,down },
  pos/n .code:n = \tl_set:Nn \l_miguel_label_pos_tl { hc,up },
  pos/w .code:n = \tl_set:Nn \l_miguel_label_pos_tl { left,vc },
  pos/s .code:n = \tl_set:Nn \l_miguel_label_pos_tl { hc,down },
  pos/e .code:n = \tl_set:Nn \l_miguel_label_pos_tl { right,vc },
  pos .initial:n = nw,
  unknown .code:n   = \clist_put_right:Nx \l_miguel_label_clist
                       { \l_keys_key_tl = \exp_not:n { #1 } }
 }
\clist_new:N \l_miguel_label_clist
\box_new:N \l_miguel_label_box
\box_new:N \l_miguel_label_image_box
\NewDocumentCommand{\xincludegraphics}{O{}m}
 {
  \group_begin:
  \tl_clear:N \l_miguel_label_tl
  \clist_clear:N \l_miguel_label_clist
  \keys_set:nn { miguel/label } { #1 }
  \tl_if_empty:NTF \l_miguel_label_tl
   {
    \miguel_includegraphics:Vn \l_miguel_label_clist { #2 }
   }
   {
    \SetHorizontalCoffin\imagecoffin
     {
      \miguel_includegraphics:Vn \l_miguel_label_clist { #2 }
     }
    \SetHorizontalCoffin\labelcoffin
     {
      \raisebox{\depth}
       {
        \bool_if:NTF \l_miguel_label_box_bool
         { \fcolorbox{white}{white}{\l_miguel_label_size_tl\l_miguel_label_tl} }
         { \l_miguel_label_size_tl\l_miguel_label_tl }
       }
     }
    \SetVerticalPole\imagecoffin{left}{3pt+\CoffinWidth\labelcoffin/2}
    \SetVerticalPole\imagecoffin{right}{\Width-3pt-\CoffinWidth\labelcoffin/2}
    \SetHorizontalPole\imagecoffin{up}{\Height-3pt-\CoffinHeight\labelcoffin/2}
    \SetHorizontalPole\imagecoffin{down}{3pt+\CoffinHeight\labelcoffin/2}
    \use:x{\JoinCoffins\imagecoffin[\l_miguel_label_pos_tl]\labelcoffin[vc,hc]} 
    \TypesetCoffin\imagecoffin
   }
   \group_end:
 }
\NewDocumentCommand{\setlabel}{m}
 {
  \keys_set:nn { miguel/label } { #1 }
 }
\cs_new_protected:Nn \miguel_includegraphics:nn
 {
  \includegraphics[#1]{#2}
 }
\cs_generate_variant:Nn \miguel_includegraphics:nn { V }
\ExplSyntaxOff

\begin{document}

\title{Absence of quantization in the circular photogalvanic effect in disordered chiral Weyl
Semimetals}
\author{Ang-Kun Wu}
\affiliation{Department of Physics and Astronomy, Center for Materials Theory, Rutgers University, Piscataway, New Jersey 08854, USA}
\affiliation{Theoretical Division, T-4 and CNLS, Los Alamos National Laboratory, Los Alamos, New Mexico 87545,USA}
\author{Daniele Guerci}
\affiliation{Center for Computational Quantum Physics, Flatiron Institute, New York, NY 10010, USA}
\author{Yixing Fu}
\affiliation{Department of Physics and Astronomy, Center for Materials Theory, Rutgers University, Piscataway, New Jersey 08854, USA}
\author{Justin H. Wilson}
\affiliation{Department of Physics and Astronomy, Louisiana State University, Baton Rouge, Louisiana 70803, USA}
\affiliation{Center for Computation and Technology, Louisiana State University, Baton Rouge, Louisiana 70803, USA}
\author{J. H. Pixley}
\affiliation{Department of Physics and Astronomy, Center for Materials Theory, Rutgers University, Piscataway, New Jersey 08854, USA}
\affiliation{Center for Computational Quantum Physics, Flatiron Institute, New York, NY 10010, USA}
\date{\today}

\begin{abstract}
The circularly polarized photogalvanic effect (CPGE) is studied in chiral Weyl semimetals with short-ranged quenched disorder. Without disorder, the topological properties of chiral Weyl semimetals lead to quantization of the CPGE, which is a second-order optical response. Using a combination of diagrammatic perturbation theory in the continuum and exact numerical calculations via the kernel polynomial method on a lattice model, we show that disorder perturbatively destabilizes the quantization of the CPGE.
\end{abstract}

\maketitle

\section{Introduction}

The prediction and discovery of gapped and gapless topological band structures changed the current perspective of quantum materials.
While gapped topological insulating band structures afford a level of protection that is provided by the nonzero electronic energy gap, gapless topological semimetals are sensitive to perturbations from interactions and disorder. 
Moreover, while gapped topological phases tend to have quantized responses, finding such quantities in gapless topological materials remains challenging. 

\begin{figure}[b!]
\begin{center}
\includegraphics[width = 0.48\textwidth]{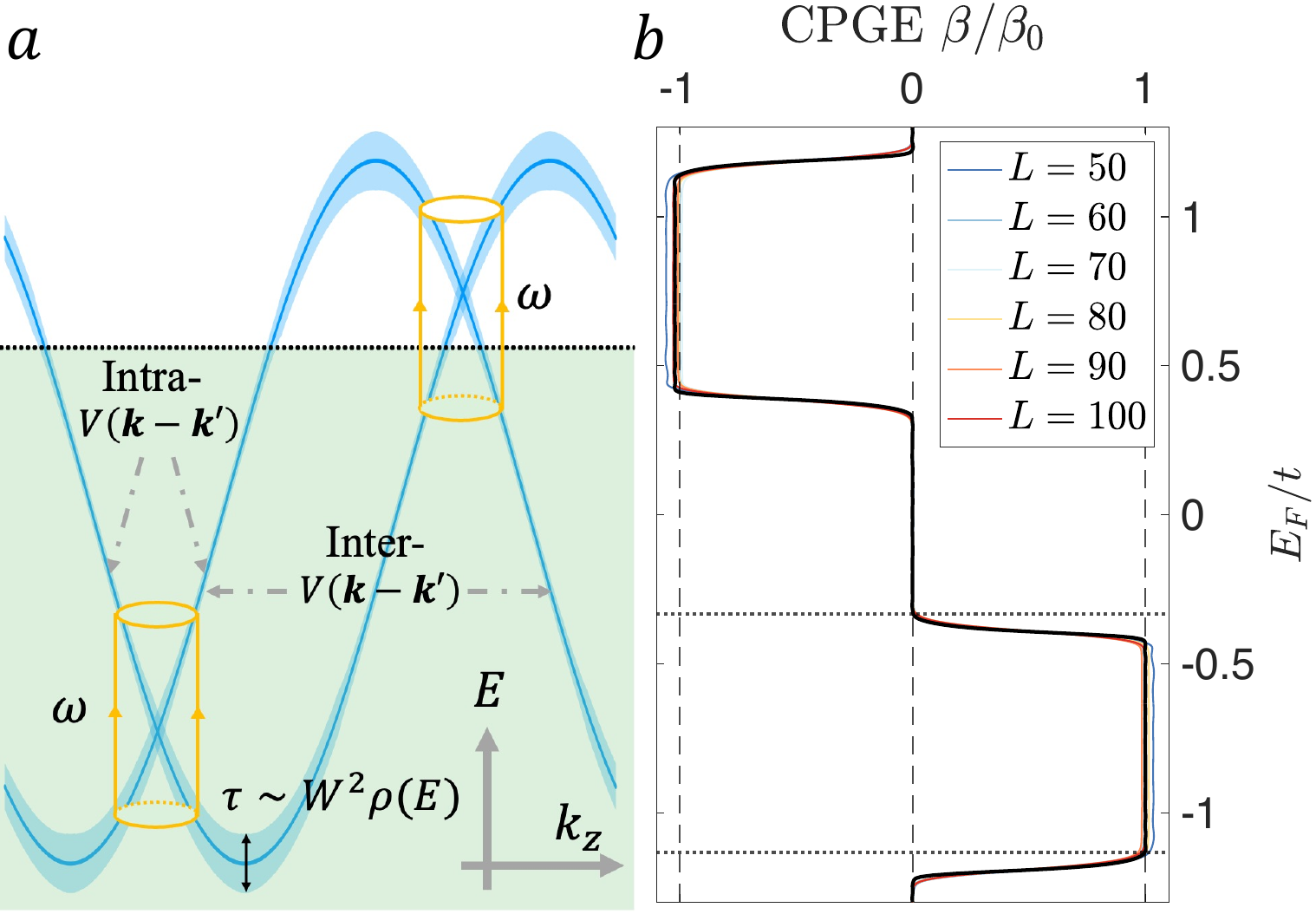}
\caption{(a) The dispersion relation as a function of $k_z$ of a two-band chiral Weyl semimetal (see \textit{Model} section). 
The onsite potential disorder $V(\mathbf{r})$ generates scattering $V(\mathbf{k-k'})$ in momentum space, giving a finite quasiparticle lifetime $\tau$. 
(b) Quantized CPGE as a function of the Fermi level with various system sizes. Vertical dashed lines mark the quantized values $\pm 1$ and $0$. Horizontal dotted lines mark two boundaries of the quantized signal and are obtained by setting $E_+(\mathbf{k})-E_-(\mathbf{k})\approx \omega$, equal to the optical frequency. The numerical setting (see Appendix \ref{App:A}): The KPM expansion order is $N_C=502$, except for the dark black curve ($L=100$ with $N_C=1024$). The half-bandwidth scale in the KPM is taken as $D=5.28t$.
}
\label{fig:dispersion}
\end{center}
\end{figure}

In chiral Weyl semimetals,  it was recently shown that   circular photogalvanic effect (CPGE), a DC photocurrent ({\bf j}) induced by  circularly polarized light at second order in the electric field
~\cite{PhysRevB.79.121302,yuan2014generation,mciver2012control,PhysRevB.93.081403,PhysRevMaterials.2.024202}, can acquire a quantized response due to separately probing the chirality of the Weyl points \cite{belinicher1980photogalvanic,asnin1979circular,ivchenko1982photogalvanic}.
Chiral Weyl semimetals have a dispersion with linearly touching points in the Brilloun zone that do not occur at the same energy due to broken mirror symmetries.
In particular, it was found that in chiral Weyl semimetals, the CPGE becomes quantized over a range of optical frequencies ($2|\mu_1|< \omega <2|\mu_2|$ assuming that there is only a pair of Weyl points that occur at $\mu_1$ and $\mu_2$, the energy differences between the two Weyl nodes to the Fermi level $E_F$, and $|\mu_2|>|\mu_1|$) constrained by positions of the Weyl nodes \cite{de2017quantized}
\begin{equation}
    \frac{\mathrm{d}\mathbf{j}}{\mathrm{d}t}=i\beta(\omega)\mathbf{E_\omega}\times\mathbf{E_{-\omega}}, \quad \beta(\omega)= \frac{\pi e^3}{3h^2}C=\beta_0C, 
\end{equation}
with the electron charge $e$, the Planck constant $h$, and $C$ is the integer-valued chirality (or topological ``charge'') of the two Weyl nodes that are exposed from Pauli blocking by breaking the mirror symmetries, see Fig.~\ref{fig:dispersion}$\mathbf{a}$. The constraint of optical frequencies lies in exciting electrons only for one Weyl node. 
This quantization can be connected to a Berry phase effect \cite{PhysRevLett.105.026805,deyo2009semiclassical,PhysRevLett.115.216806,PhysRevB.83.035309}.
However, experiments on the putative chiral Weyl material RhSi did not observe a clear quantization of the CPGE and its expected quantized value seems to depend partially on the finite lifetime of photoexcited carriers, as it was used to  estimate the expected quantized value~\cite{rees2020helicity,ni2021giant,ni2020linear}.

Strikingly, generic electron-electron interactions in a chiral Weyl semimetal can break the perfect quantization of the CPGE \cite{PhysRevLett.124.196603,sym12060919}, in contrast to the robustness of the quantum Hall effect \cite{PhysRevLett.54.259,PhysRevB.31.3372,ishikawa1986magnetic,hastings2015quantization,coleman1985no,PhysRevB.94.205139,giuliani2017universality} or the chiral anomaly \cite{PhysRev.177.2426,PhysRev.182.1517,giuliani2021anomaly,PhysRevB.86.115133,burkov2015chiral,burkov2018weyl}. 
While RhSi is a weakly correlated material, its interaction effects could be either much less than, or comparable to, the effects of disorder scattering depending on the amount of screening in the material. 
Also, it has been established previously that weak impurity scattering introduces several scattering channels that can disrupt the CPGE quantization~\cite{PhysRevB.96.075123}.
As such, it is essential to develop a theoretical understanding of the effects of disorder on the quantization of the CPGE in chiral Weyl semimetals.

In this letter, we demonstrate that short-range quenched disorder \cite{PhysRevB.87.155123,roy2016universal} destroys the quantization of the CPGE in chiral Weyl semimetals. This is demonstrated in a lattice model through a combination of diagrammatic perturbation theory and exact numerical calculations based on the kernel polynomial method (KPM) for higher-order response functions. We show that due to the finite density of states at the Fermi energy, perturbatively generated disorder scattering is sufficient to generate a finite quasiparticle lifetime ($\tau$) and shift of the excitation spectrum, which coalesce to produce a correction to the height of the quantized plateau to decrease to leading order in the disorder strength ($W$) as $\sim W^2$. Going beyond perturbation theory, we utilize the KPM implemented on GPUs to efficiently compute a triple Chebyshev expansion of the CPGE allowing us access to large system sizes that definitively shows that the CPGE is not quantized in the presence of disorder and that perturbation theory is sufficient in the weak disorder regime. While non-perturbative effects of the random potential are present for a weak mirror symmetry breaking term, these produce exponentially small corrections (in the disorder strength), which are subleading relative to the polynomial corrections from perturbation theory.

\section{Model and Approach}

To emulate the chiral Weyl semimetal without mirror symmetry, we consider a two-band model whose Hamiltonian can be conveniently written in the momentum space
$
        H_\mathbf{k}= \mathbf{d}_\mathbf{k} \cdot \boldsymbol{\sigma} +\gamma \sin(k_z) \sigma^0
$
where $\boldsymbol{\sigma}=\{\sigma^x,\sigma^y,\sigma^z\}$ is the vector of Pauli matrices and
$d^x_\mathbf{k}=-t\sin(k_x), d^y_\mathbf{k}=-t\sin(k_y)$ and $d^z_\mathbf{k}=-M+t\sum_{i=x,y,z}\cos(k_i)$ with $M=2t$. The second term is diagonal $\sigma^0=I_{2\times 2}$, which breaks the mirror symmetry and shifts the energy of the two Weyl nodes to energy $E = \pm \gamma$ at positions $\mathbf{k}=\{0,0,\pm \pi/2 \}$. In real space, the model can be written as
\begin{equation*}
    \begin{split}
        H_0 &=  \sum_\mathbf{r}\bigg\{ -\psi_\mathbf{r}^\dagger M \sigma^z \psi_\mathbf{r} +\sum_{\hat{\boldsymbol{\alpha}}} [\psi_{\mathbf{r}+\hat{\boldsymbol{\alpha}}}^\dagger M_{\hat{\boldsymbol{\alpha}}} \psi_{\mathbf{r}}+\mathrm{h.c.}]\bigg \},
    \end{split}
\end{equation*}
with $\psi_\mathbf{r}=(a_\mathbf{r},b_\mathbf{r})^T$ a two-component Pauli spinor and $\hat{\boldsymbol{\alpha}}=\hat{\mathbf{x}},\hat{\mathbf{y}},\hat{\mathbf{z}}$. The hopping matrices can be expressed via Pauli matrices $M_{\hat{\mathbf{x}}}=\frac{t}{2}(\sigma^z+i\sigma^x)$, $M_{\hat{\mathbf{y}}}=\frac{t}{2}(\sigma^z+i\sigma^y)$ and $M_{\hat{\mathbf{z}}}=\frac{t}{2}\sigma^z-\frac{i\gamma}{2}\sigma^0$. The dispersion of this two-band chiral Weyl semimetal is shown in Fig.~\ref{fig:dispersion}$\mathbf{a}$. Adding short-range onsite potential disorder to the system gives
\begin{equation}
    H = H_0+\sum_\mathbf{r} \psi^\dagger_\mathbf{r} V(\mathbf{r}) \psi_\mathbf{r},
    \label{eqn:H}
\end{equation}
where $V(\mathbf{r})$ are local onsite potentials drawn independently from the Gaussian distribution with zero mean and standard deviation $W$.
To avoid the finite-size effect introduced by the finite total disorder energies $E_0=\frac{1}{L^3}\sum_\mathbf{r}V(\mathbf{r})$ due to randomness, the potential is shifted accordingly, $V_i(\mathbf{r})\to V_i(\mathbf{r})-E_0$.

The disorder potential in the model breaks the translational symmetry of the system, and therefore we evaluate the CPGE in the energy eigenbasis. The
second-order photocurrent (in units of $\hbar=1$) is
\begin{equation}
    \begin{split}
        J^\gamma(\Omega)=\frac{\chi^{\alpha\beta\gamma}(\omega_1,\omega_2)+\chi^{\beta\alpha\gamma}(\omega_2,\omega_1)}{\omega_1\omega_2}\varepsilon_{\alpha\beta\gamma}E^\alpha_{\omega_1} E^\beta_{\omega_2},
    \end{split}
\end{equation}
where $\varepsilon_{\alpha\beta\gamma}$ is the Levi-Civita symbol and $\Omega=\omega_1+\omega_2$. Circularly polarized light at a specific frequency is ideal for observing the frequency limit $\Omega = 0$ (the DC limit). In this limit, this second-order optical response
\begin{equation}
    y(\omega_1,\omega_2)\equiv \mathrm{Im}\bigg[\frac{\chi^{\alpha\beta\gamma}(\omega_1,\omega_2)+\chi^{\beta\alpha\gamma}(\omega_2,\omega_1)}{\omega_1\omega_2}\bigg]
    \label{eqn:y}
\end{equation}
reduces to the CPGE response 
\begin{equation}\label{eq:beta}
    \beta(\omega)\equiv \lim_{\Omega\to 0} \Omega y(\omega,\Omega-\omega),
\end{equation}
which captures the quantization \cite{PhysRevLett.124.196603}. Note that the bare response $y(\omega,\Omega-\omega)$ diverges at quantization in the limit $y(\omega,\Omega-\omega)\sim \frac{\beta_0 C}{\Omega}$ for clean systems while for finite size or energy resolution, the divergence in $y(\omega,-\omega)$ is rounded out by a scale that is proportional to the broadening of the Green's function as discussed below in Eq.~\eqref{eq:KPMreg}.
The basis independent expression for the second-order response is given by \cite{joao2019basis} 
\begin{widetext}
\begin{equation}\label{eq:chi}
\begin{split}
        \chi^{\alpha\beta\gamma}(\omega_1,\omega_2)
        &=\frac{e^3}{V\hbar^2}\int \mathrm{d}\epsilon f(\epsilon)\mathrm{Tr}\bigg[\hat{j}^\alpha G^R(\epsilon/\hbar+\Omega) \hat{j}^\beta G^R(\epsilon/\hbar+\omega_2)\hat{j}^\gamma\delta(\epsilon-H)\\
        &+\hat{j}^\alpha G^R(\epsilon/\hbar+\omega_1)\hat{j}^\beta \delta(\epsilon-H)\hat{j}^\gamma G^A(\epsilon/\hbar-\omega_2)+\hat{j}^\alpha \delta(\epsilon-H) \hat{j}^\beta G^A(\epsilon/\hbar-\omega_1)\hat{j}^\gamma G^A(\epsilon/\hbar-\Omega)\bigg],
\end{split}
\end{equation}
\end{widetext}
where $\hat{j}^\alpha,\hat{j}^\beta,\hat{j}^\gamma$ are current operators at $\alpha,\beta,\gamma=x,y,z$ direction and $G^{R/A}(\omega)$ stands for  retarded and advanced Green's function corresponding to $\omega \pm i\eta$ for $\eta\rightarrow 0^+$,
 respectively. $V=V_cN$ is the total volume, where $V_c$ is the volume of one unit cell (taken as $1$ and set $\hbar=e=1$) and $N=L^3$ is the total number of unit cells. $f(\epsilon)$ is the Fermi distribution, and we focus on $T=0$ case. With periodic boundary condition (PBC), the current operator can be obtained by multiplying the Hamiltonian by the position difference operator $\hat{j}^\alpha=(i\hbar)^{-1} Hd^\alpha,$ with $d^\alpha_{ij}$ denoting the position difference of the atom $i,j$ in the $\alpha$ direction \cite{joao2019basis}. The quantity $\chi^{\alpha\beta\gamma}$ is imaginary due to these three current operators. In Eq.~\eqref{eq:chi}, for every chosen $\omega_1=\omega$, we integrate up to the Fermi level $E_F$. However, the quantization condition $2|\mu_1|< \omega <2|\mu_2|$ can be conveniently changed into $2|\gamma-\lvert E_F\rvert| < \omega <2|\gamma+\lvert E_F\rvert |$ 
where $\pm\gamma$ are positions of Weyl nodes in energy relative to $E_F=0$. When we fix the optical frequency $\omega$ and vary $E_F$, we only need to compute and integrate the Green's functions (depending both on $E_F,\omega$) once, which is computationally convenient and does not change the quantization physics.

In order to reach large system sizes, we employ the kernel polynomial method (KPM) and expand the delta function and Green's functions in the response $\chi^{\alpha\beta\gamma}$ in terms of Chebyshev polynomials. Based on the work from \cite{joao2019basis}, we find the KPM estimate for  this part of the second-order conductivity to be
\begin{equation}
    \begin{split}
        &\chi_{\mathrm{KPM}}^{\alpha\beta\gamma}(\omega_1,\omega_2)=\frac{e^3}{\hbar^3}\sum_{nmp}^{N_C^3} \Lambda_{nmp}(\omega_1,\omega_2)\Gamma^{\alpha,\beta,\gamma}_{nmp},\\
        &\Gamma^{\alpha,\beta,\gamma}_{nmp}=\frac{\mathrm{Tr}}{N}\bigg[\hat{j}^\alpha \frac{T_n(\tilde{H})}{1+\delta_{n0}}\hat{j}^\beta \frac{T_m(\tilde{H})}{1+\delta_{m0}}\hat{j}^\gamma \frac{T_p(\tilde{H})}{1+\delta_{p0}}\bigg],
    \end{split}
    \label{eqn:kpm}
\end{equation}
where $\Lambda_{nmp}$ is a tensor defined in the Appendix \ref{App:A}, $\delta_{ij}$ here is the Kronecker delta, and the KPM kernel is absorbed into the Dirac delta and Green's function (Appendix \ref{App:A}). $\tilde{H}=(H-b)/D$ the rescaled Hamiltonian for the KPM calculation, where $b$ and $D$ are the band asymmetry and bandwidth respectively [chosen so that $\tilde H$ has eigenvalues strictly within the interval $(-1,1)$], and we denote the KPM expansion order as $N_C$.
Note that the KPM tensor $\Gamma^{\alpha,\beta,\gamma}_{nmp}$ is averaged over $50$ disorder realizations instead of the CPGE response, while these are formally equivalent (Appendix \ref{App:A}), this allows for a computational advantage. 

Numerically, it is inevitable to introduce finite broadening in the Green's functions for finite systems. In KPM, the finite expansion order $N_C$ broadens eigenstates by $\eta_\mathrm{KPM}\approx\pi/N_C$ (Jackson kernel near the band center \cite{RevModPhys.78.275}). 
And thus any divergent feature requiring eigenstates connected by frequency $\omega$ will see responses due to all eigenstates $O(N_C^{-1})$ away in energy. Since Eq.~\eqref{eq:chi} contains integrals of Green's functions, the broadening effect in the non-linear response is nontrivial, especially in the diverging DC limit in Eq.~\eqref{eq:beta}.  Empirically,  the regularization of the divergence at the quantized response is found to scale linearly with $N_C$ under the Jackson kernel used in Eq.~\eqref{eqn:kpm} (see Appendix \ref{App:A}). Formally, we find
\begin{equation}\label{eq:KPMreg}
    \beta(\omega)=\mathcal{N}_{\mathrm{KPM}}\frac{y(\omega,-\omega)}{N_C},
\end{equation}
where $\mathcal{N}_\mathrm{KPM}\approx31.74\beta_0/D^3$ is the KPM normalization factor in clean systems. 
The proper normalization of the divergence helps us determine the effect of disorder on the quantization of the CPGE.

We focus on the Fermi energy ($E_F$) dependence while fixing the photon frequency 
for the computational advantage we mentioned earlier.
The numerical CPGE results of a clean chiral Weyl semimetal can be found in Fig.~\ref{fig:dispersion}$\mathbf{b}$. Quantization of the CPGE can be found as the plateau over a range of Fermi energies when we fix the photon frequency $\omega\equiv \omega_1=\gamma$ to have relatively long plateau signals. The width of this plateau is controlled by the energy levels at which there exists an exact excitation of electrons at a given $\mathbf{k}$ whose energy difference $\Delta E=\omega$ between the two bands. The horizontal dotted lines in Fig.~\ref{fig:dispersion}$\mathbf{b}$ are obtained by varying $k_z$ when $k_x,k_y=0$, which mark two boundaries of the plateau. The broadening of the boundaries can be found by varying all three momentum components.

\begin{figure}[t!]
\begin{center}
\includegraphics[width = 0.48\textwidth]{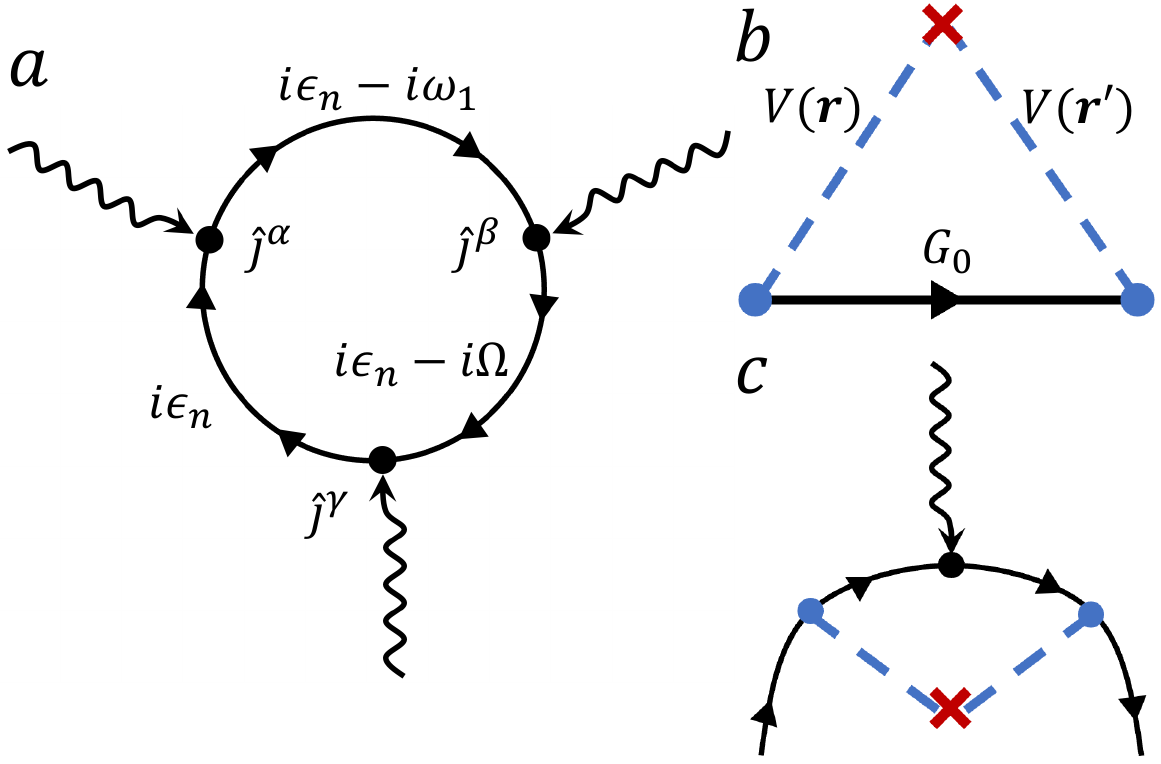}
\caption{The Feymann diagrams for perturbation theory. (a) The diagramatic form of the CPGE expression. (b) Disorder self-energy in the non-crossing approximation \cite{PhysRevResearch.3.033137}. The blue symbols represent disorder potential while the red cross is the delta function $\delta(\mathbf{r-r'})$ from disorder average. (c) The vertex diagram for one current operator (see details in Appendix \ref{SI:perturbation}).
}
\label{fig:diagrams}
\end{center}
\end{figure}

\section{Perturbative description of the CPGE}

To describe the breakdown of the CPGE quantization analytically, 
we apply diagrammatic perturbation theory in the disorder strength ($W$) to evaluate the CPGE to leading order.  
The non-perturbative rare region effects of the random potential~\cite{pixley2021rare} we will show are subleading and will be discussed in the end.
In the clean limit,
we  take the effective low-energy continuum approximation to the lattice model in Eq.~\eqref{eqn:H}
\begin{equation}
    \tilde{H}_0=\sum_{\mathbf{k},k<\Lambda_k} \sum_{a=1}^2\psi_{a\mathbf{k}}^\dagger [(-1)^{a+1}v_F\mathbf{k}\cdot \boldsymbol{\sigma}-\mu_a]\psi_{a\mathbf{k}},
\end{equation}
where $a=1,2$ denotes the two nodes at $k_{za}=(-1)^{a+1}\pi/2$ with opposite chirality, and $\Lambda_k$ is the momentum cutoff. In the lattice model, $\gamma$ controls the chemical potentials $\mu_a=E_F+(-1)^a\gamma$ at each node and the Fermi velocity is just $v_F=t$, while the momentum cutoff $\Lambda_k$ is chosen numerically for the perturbation. The single-particle Green's functions $G^{(a)}(i\varepsilon_n,\mathbf{k})$ can be analytically expressed in the Matsubara basis $\varepsilon_n$. 

\begin{figure}[t!]
\begin{center}
\includegraphics[width = 0.48\textwidth]{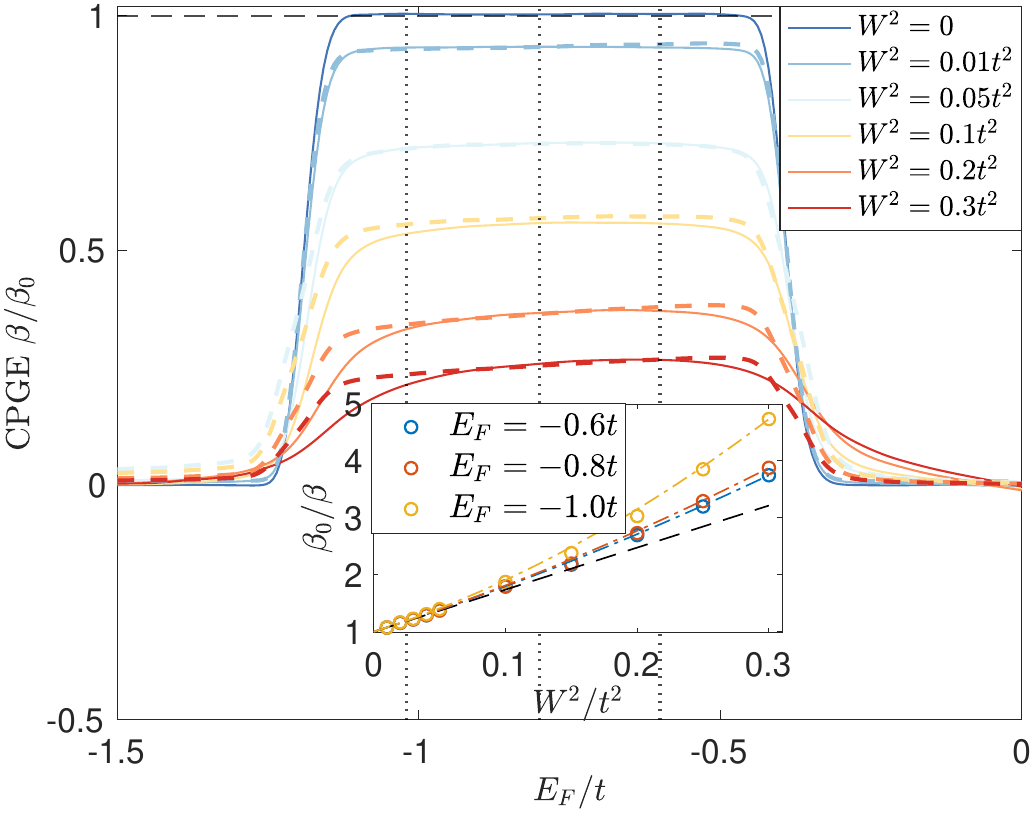}
\caption{The CPGE quantization in the clean limit and its suppression with increasing disorder strength with $L=100, N_C=512$ (solid lines). In KPM, the half-bandwidth is chosen as $D=6t$, slightly larger than the clean system half-bandwidth for sample average. 
The normalization factor is the same as in the clean case $C_\mathrm{KPM}=31.74\beta_0$. Dashed curves represent perturbative corrections with quasiparticle lifetime, frequency shift, and vertex correction. The vertical dotted lines mark the locations $E_F=[-1.02,-0.80,-0.60]t$. 
Inset: Inverse of the normalized CPGE response $\beta_0/\beta$ as a function of disorder strength $W^2$ at various Fermi levels with $N_C=512$ from marked $E_F$. The dashed curve shows a linear fit to the first $5$ small $W$ of the middle curve, where $\beta_0/\beta \approx 0.999+7.39 W^2$. The dashed-dotted curves are the second-order fit with coefficients $5.91, 7.37, 16.9$ for $W^4$ terms.
}
\label{fig:Disorder}
\end{center}
\end{figure}

The relevant diagrams to the three current operators are shown in Fig.~\ref{fig:diagrams}$\mathbf{a}$.
The first-order terms in $W$ vanish upon disorder average and the non-vanishing scattering vertex  is given in Fig.~\ref{fig:diagrams}$\mathbf{b}$, which gives rise to the self-energy correction to each single Green's functions and vertex corrections to each current vertex (Fig.~\ref{fig:diagrams}$\mathbf{c}$). Integrating up to the momentum cutoff $\Lambda$ gives 
\begin{equation}
    \begin{split}
\Sigma(i\epsilon)&=  W^2 \sum_{a=1}^{2}F_a(i\epsilon).\\
F_a(i\epsilon)&
=\rho(\mu_a)\left[-\frac{\Lambda v_{F} }{\mu_a}+\text{tanh}^{-1}\frac{v_{F}\Lambda}{\mu_a}\right]\\
&-\frac{i\pi}{2}\text{sign}(\epsilon)\rho(\mu_a),
    \end{split}
    \label{eqn:selfenergy}
\end{equation}
where the clean, low-energy density of states is $\rho(\epsilon)=\frac{3\epsilon^2}{2(v_F\Lambda)^{3}}$, 
and $\epsilon \in[-v_F\Lambda,v_F\Lambda]$. The imaginary part of $F_a(i\epsilon)$ gives a finite quasiparticle lifetime $\tau_W=(2\eta_W)^{-1},\eta_W =-\mathrm{Im}(\Sigma)\propto W^2$ \cite{altland2010condensed}, while the real part gives corrections that approximately shift the frequency $\omega\to \omega-\mathrm{Re}(\Sigma)$ where the quantized plateau sets in. Besides this self energy, we also perturbatively compute the leading vertex corrections (see Appendix \ref{SI:perturbation}) such that
the CPGE is given by
\begin{equation}
    \delta \beta(E_F) =-\beta_0(E_F)\frac{4W^2}{\pi^2v_F^4\Lambda}[\mu_1^2-\mu_2^2+\omega_1(\mu_2-\mu_1)],
    \label{eqn:pertCPGE}
\end{equation}
where the Fermi energy dependency hides in $\mu_a(E_F)$ and $\beta_0(E_F)$ represents the CPGE from a clean system
(Appendix \ref{SI:perturbation}), thus the plateau value of the CPGE
decreases like $\sim -W^2$. 
The perturbative results for each disorder strength are also shown as dashed curves in Fig.~\ref{fig:Disorder}, which fit well only in the weak disorder limit ($W^2\lesssim 0.05t^2$) as we are only considering the leading perturbative contribution in the low-energy limit
and this therefore will break down as $W$ increases due to higher-order corrections and band curvature effects, both of which are captured in the numerics. Therefore, we now turn to a numerically exact evaluation of the disorder-averaged CPGE and use the perturbative results to provide a physical interpretation and grounding to the numerical results.

\section{Numerical evaluation of the CPGE}

We evaluate $\mathrm{Im}(\chi)$ in the triple KPM expansion in Eq.~\eqref{eqn:kpm} using GPUs to obtain the CPGE response in Eq.~\eqref{eq:beta} in the $\Omega=0$ limit. In all results presented below, we regularize the divergence of the response  using the KPM based scaling in Eq.~\eqref{eq:KPMreg} to extract the $N_C$ independent response, and we take the system size $L$ sufficiently large (see Appendix \ref{App:A}) so that our estimate is independent of the finite size. This provides an estimate of the CPGE response in the thermodynamic limit that is free from rounding out the response from finite size and energy resolution.
As shown in Fig.~\ref{fig:Disorder}, the CPGE responses are presented for various disorder strengths. Quantized responses in the clean limit decrease continuously with increasing disorder strength, which agrees quantitatively well with perturbation theory in small $W$.
In addition to the reduction in plateau height, we also observe that the responses develop a linear slope at small $W^2$ and quadratic curvature at large $W^2$ (inset of Fig.~\ref{fig:Disorder}) and a clear broadening around the edges of the plateau (Appendix \ref{App:C}), which are due to a disorder shift in the excitation spectrum and the finite quasiparticle lifetime, respectively.

\begin{figure}[t!]
\begin{center}
\includegraphics[width = 0.48\textwidth]{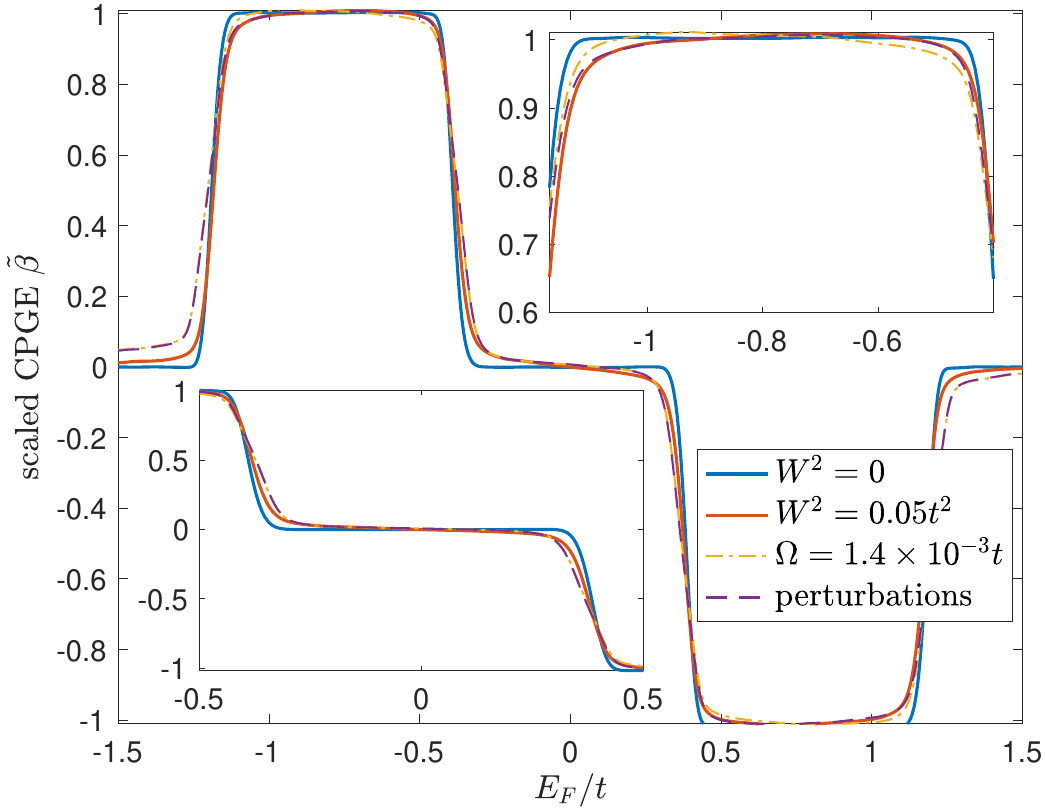}
\caption{The perturbative deformation to the CPGE quantization. The $y-$axis $\tilde{\beta}$ is rescaled to $[-1,1]$ to compare different results (while also effectively removing the contribution from the quasiparticle lifetime)
The system size $L=100$ with KPM expansion order $N_C=512$. The clean results shifted to a finite $\Omega_{\Sigma}$ to simulate the effect of the real part of the self energy in Eq.~\eqref{eqn:selfenergy}
are obtained by setting $\omega_1'=\omega_1+\Omega, \omega_2=-\omega_1$ with a clean and periodic system. 
Inset: The zoomed-in region around $E_F=0,0.8t$ to show the plateau deformation from vertex corrections.
}
\label{fig:correction}
\end{center}
\end{figure}

Fixing the Fermi energies at the marked locations, the disorder dependence of the bare response $y$ can be found in the inset of Fig.~\ref{fig:Disorder}. We find that it is well described by the perturbative expression
\begin{equation}
    \beta \approx \beta^{(0)} + \frac{\beta^{(2)}}{2!} W^2+ \frac{\beta^{(4)}}{4!} W^4+\dots 
\end{equation}
where $\beta^{(n)}$ are $n$th derivative of the CPGE with respect to $W$ in the clean limit. At plateau range, $\beta^{(2)}<0$ while $\beta^{(2)}>0$ at plateau edges (SI Fig.~5). After extracting the $N_C$ independent response, we find that  $\beta_0/\beta \approx 1+C_WW^2$ (Fig.~\ref{fig:Disorder} inset, $C_W$ is the coefficient for disorder), which leads to the conclusion that the plateau value is set by the disorder strength as $\beta/\beta_0 \approx 1-C_W W^2$ and is not quantized in the small-$W$ limit (Appendix \ref{App:C}). 
This can be further approximated to match the perturbative expression in Eq.~\eqref{eqn:pertCPGE} at sufficiently small $W$, where the momentum cutoff $\Lambda\approx 0.52t$ from numerical fitting while $\Lambda\approx 0.48t$ from the dispersion \footnote{In Fig.~\ref{fig:dispersion} $\mathbf{a}$, an effective momentum cutoff can be approximated by measuring the energy difference from one Weyl node to its bended edge}. The coefficients depend on the energy at which it is evaluated, as depicted in Fig.~\ref{fig:correction}. Physically, we can interpret the dependence of the plateau height being dominated by the quasiparticle lifetime, which perturbatively admits a polynomial expansion $1/\tau(E_F)=a_0W^2+a_1 W^4 +\dots$. On the other hand, the strong deformation of the plateau edges 
is instead due to the shift of the excitation spectrum from disorder scattering, which is given by the real part of the self-energy in Eq. \eqref{eqn:selfenergy}.

To understand each contribution to the CPGE
in Fig.~\ref{fig:correction} we compare the disorder-averaged CPGE (red line) with the clean result (blue curve) and the clean result shifted to a finite frequency $\Omega_{\Sigma}$ (yellow curve) that represents the contribution from the real part of the self energy, and we rescale the height to account for the imaginary part of the self energy. Therefore, we are able to analyze the vertex corrections on top of the self-energy by comparing the clean result with a frequency shift $\Omega_{\Sigma}$ with the leading perturbative result in Eq.~\eqref{eqn:pertCPGE} contributing mainly to the deformation of the flatness of the plateau (see inset of Fig.~\ref{fig:correction}). Note that $\Omega_\Sigma$ is empirical and also deforms the flatness but in such a way that the vertex corrections dominates in the disordered results.

In the Appendix \ref{App:D}, we discuss the non-perturbative effects of rare regions of random potentials \cite{pixley2021rare} that survive the mirror symmetry breaking perturbation $\gamma \lesssim 0.3$, which is a sub-leading exponentially small (in disorder strength correction) correction to the $W^2$ contribution and is hence  indistinguishable from having only perturbative states in the spectrum. Thus, the destabilization of the CPGE is a perturbative effect arising from the metallic nature of chiral Weyl semimetals.

\section{Conclusions}

In this work, we have shown that adding disorder to a chiral Weyl semimetal destroys the CPGE quantization both numerically on finite-size systems and perturbatively, which agree well in the weak disorder regime. Taken with Ref.~\cite{PhysRevLett.124.196603}, we conclude that the CPGE does not remain quantized in the presence of disorder or interactions and will therefore not be quantized in any material realization. 

\section*{Acknowledgement}

We thank Daniel Kaplan, Jo\~ao Manuel Viana Parente Lopes, Joel Moore, and Peter Orth for useful discussions. This work is partially supported by NSF Career Grant No.~DMR- 1941569 and the Alfred P.~Sloan Foundation through a Sloan Research Fellowship (A.K.W., Y.F., J.H.P.), as well as NSF CAREER Grant No.~DMR-2238895 (J.H.W.). Part of this work was performed in part at the Aspen Center for Physics, which is supported by the National Science Foundation Grant No.~PHY-2210452 (J.H.W., J.H.P.) as well as the Kavli Institute of Theoretical Physics that is supported in part by the National Science Foundation under Grants No.~NSF PHY-1748958 and PHY-2309135 (J.H.W., J.H.P.). A.K.W. is also partially supported by the US DOE NNSA under Contract No. 89233218CNA000001 through the LDRD Program and was performed, in part, at the Center for Integrated Nanotechnologies, an Office of Science User Facility operated for the U.S. DOE Office of Science, under user proposals \#2018BU0010 and \#2018BU0083.

\appendix

\section{KPM formalism for the CPGE}\label{App:A}

To reach large system sizes, we employ the kernel polynomial method (KPM) and expand the delta function and Green's functions in the response $\chi^{\alpha\beta\gamma}$ in terms of Chebyshev polynomials. Based on the work from \cite{joao2019basis,PhysRevLett.114.116602}, we find the expression is part of the second-order conductivity
\begin{equation}\label{eq:CPGEKPM}
    \begin{split}
        &\chi^{\alpha\beta\gamma}(\omega_1,\omega_2)=\frac{e^3}{\hbar^3}\sum_{nmp} \Lambda_{nmp}(\omega_1,\omega_2)\Gamma^{\alpha,\beta,\gamma}_{nmp},
    \end{split}
\end{equation}
\begin{equation}
    \begin{split}
        &\Gamma^{\alpha,\beta,\gamma}_{nmp}=\frac{\mathrm{Tr}}{N}\bigg[\hat{j}^\alpha \frac{T_n(\tilde{H})}{1+\delta_{n0}}\hat{j}^\beta \frac{T_m(\tilde{H})}{1+\delta_{m0}}\hat{j}^\gamma \frac{T_p(\tilde{H})}{1+\delta_{p0}}\bigg],
    \end{split}
\end{equation}
where $\tilde{H}=(H-b)/D$ the rescaled Hamiltonian for the KPM calculation, and $\delta_{ij}$ here is the Kronecker delta (different from the Dirac deltas of the system defined below).
The function $\Lambda_{nmp}$ is a numerically tensor
\begin{widetext}
\begin{equation}
    \begin{split}
        \Lambda_{nmp}(\omega_1,\omega_2)&=\hbar^2\int_{-\infty}^\infty \mathrm{d}\epsilon f(\epsilon)\bigg[g_n^R(\epsilon/\hbar+\omega_1+\omega_2)g_m^R(\epsilon/\hbar+\omega_2)\Delta_p(\epsilon)\\
        &+g_n^R(\epsilon/\hbar+\omega_1)\Delta_m(\epsilon)g_p^A(\epsilon/\hbar-\omega_2)+\Delta_n(\epsilon)g_m^A(\epsilon/\hbar-\omega_1)g_p^A(\epsilon/\hbar-\omega_1-\omega_2)\bigg]
    \end{split}
\end{equation}
\end{widetext}
where the Jackson kernel $K_n^J$ is absorbed into the delta and Green's function
\begin{equation}
    \begin{split}
        \Delta_n(\epsilon)&=\frac{2T_n(\epsilon)}{\pi\sqrt{1-\epsilon^2}}K_n^J,\\
        g_n^{\pm,0^+}(\epsilon)&=\mp 2i\frac{e^{\mp i n \arccos (\epsilon\pm i 0^+)}}{\sqrt{1-(\epsilon\pm i 0^+)^2}}K_n^J\\
        K_n^J&=\frac{1}{N_C+1}\bigg[(N-n+1)\cos \frac{\pi n}{N_C+1}\\
        &+\sin\frac{\pi n}{N_C+1}\cot\frac{\pi}{N_C+1}\bigg], 
    \end{split}
\end{equation}
and $T_n(x)=\cos(n\arccos (x))$ is the first kind of the Chebyshev polynomials. Here, $g_n^{+,0^+}$ represents the retarted Green's function while  $g_n^{-,0^+}$ is the advanced one. In deriving the Eq. \ref{eq:CPGEKPM}, the Dirac deltas and Green’s functions are implicitly expanded in terms of Chebyshev polynomials
\begin{equation}
    \begin{split}
        \delta(\epsilon-\tilde{H})&=\sum_{n=0}^{N_C-1}K_n^J \Delta_n(\epsilon)\frac{T_n(\tilde{H})}{1+\delta_{n0}},\\
        g^{\pm,0^+}(\epsilon,\tilde{H})&=\frac{\hbar}{\epsilon-\tilde{H}\pm i0^+}\\
        &=\hbar \sum_{n=0}^{N_C-1} K_n^J g_n^{\pm,0^+}(\epsilon) \frac{T_n(\tilde{H})}{1+\delta_{n0}}.
    \end{split}
\end{equation}

\begin{figure}[t!]
\begin{center}
\includegraphics[width = 0.49\textwidth]{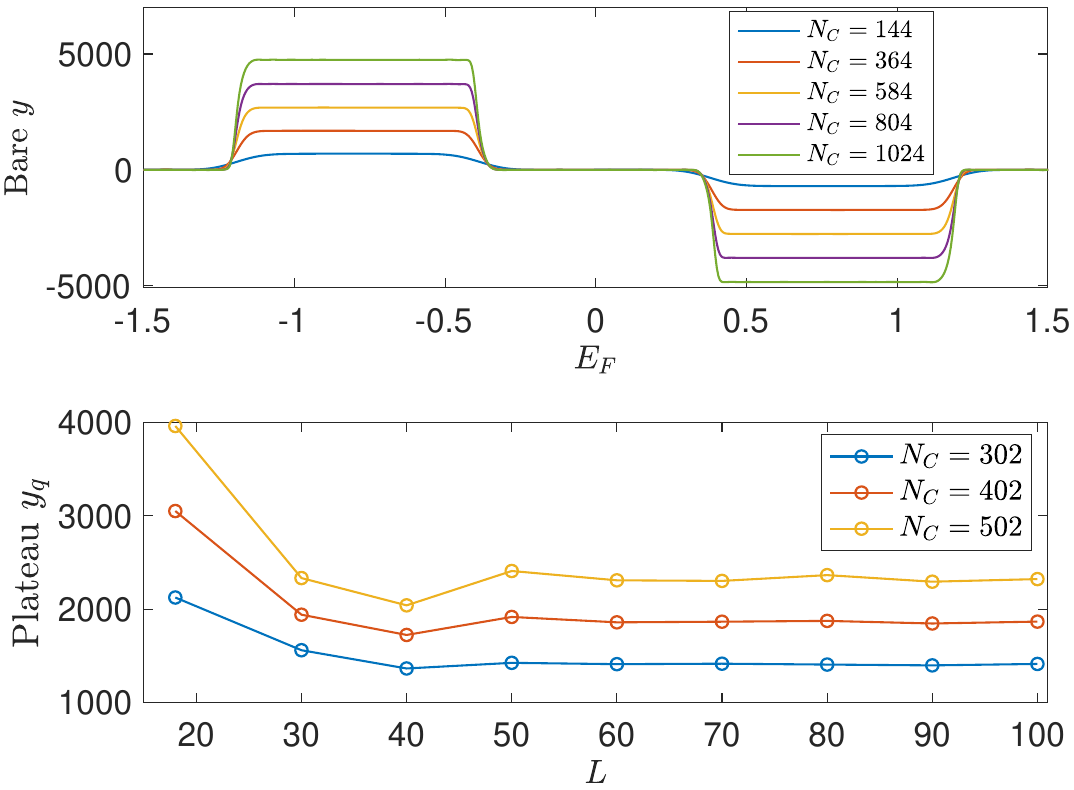}
\caption{The bare response of the CPGE with various KPM expansions $N_C$ and system sizes $L$. Clean Weyl systems are simulated with the periodic boundary condition.  The bare value of $\Omega$ is taken to be exactly 0. We take $\omega_1=\gamma=0.8t$ to have relatively large plateau. Upper panel: Bare response as a function of $E_F$ with $L=100$. The quantization is manifested by the clear plateau while the heights increase with $N_C$. Lower panel: The height of plateau as a function of system sizes. For fixed $N_C$, quantization is converged with large $L$. }
\label{fig:NCscale}
\end{center}
\end{figure}

 From the main text, Eq. (5), we know that the CPGE response is diverging in the DC $\Omega\to 0$ and thermodynamic limit $L\to \infty$. Taking $\Omega=0$ we have the KPM estimate $\beta(E_F)=\mathcal{N}y(E_F)$ ($\mathcal{N}$ some normalization) where the quantized plateau value of $y(E_F)$, denoted $y_q$ will be strongly dependent on $N_C$. For sufficiently large system sizes $\mathcal{N}=\mathcal{N}_{\mathrm{KPM}}/N_C$ is $L$ independent and we are able to find $\mathcal{N}_{\mathrm{KPM}}=31.74 \beta_0/D^3$ where $D$ is the full bandwidth, as shown in Fig. \ref{fig:NCscale}. As a result, the height of the plateau in the clean limit with KPM goes as $y_q = N_C \beta_0/\mathcal{N}_{\mathrm{KPM}}$. Note that for convenience, the default energy scale is all in terms of $t$, the hopping strength of the two band model, i.e., $E_F\equiv E_F/t, W\equiv W/t$.

Note that in the main text, results are shown without error bars due to the averaging over the KPM tensor $\Gamma_{nmp}^{\alpha,\beta,\gamma}$
\begin{equation}
    \langle \beta\rangle \sim \langle \chi^{\alpha\beta\gamma}\rangle \sim \langle \Gamma_{nmp}^{\alpha,\beta,\gamma}\rangle,
\end{equation}
where $\langle \cdots \rangle$ represents disorder average, $\sim$ means linearly dependent and $\Lambda_{nmp}(\omega_1,\omega_2)$ is system-independent integrals as we fixed $\omega_1,\omega_2$. By averaging $\Gamma_{nmp}^{\alpha,\beta,\gamma}$, we do not need to compute the triple sum $\mathcal{O}(N_C^3)$ for $\chi^{\alpha\beta\gamma}$, and thus save a lot of computational effort. The disorder-sample-dependent effect is small and discussed with the rare region effects in the last section. Also, in order to conduct the KPM tensor averaging, we fix the half-bandwidth $D$ for all disorder samples (instead of varying $D$ for every sample, and $b=0$ in $\tilde{H}=(H-b)/D$) to keep the energy in the same scale, where we sacrifice a little KPM resolution $\mathcal{O}(D/N_C)$ by slightly larger $D$ for computational advantage.

\section{Perturbation Theory for the CPGE from effective low-energy model}\label{SI:perturbation}

Consider the effective low-energy continuum theory,
\begin{equation}
    H_0=\sum_\mathbf{k} \bigg[\psi_{1\mathbf{k}}^\dagger (v_F\mathbf{k}\cdot \boldsymbol{\sigma}-\mu_1)\psi_{1\mathbf{k}}+\psi_{2\mathbf{k}}^\dagger (-v_F\mathbf{k}\cdot \boldsymbol{\sigma}-\mu_2)\psi_{2\mathbf{k}}\bigg],
\end{equation}
with $v_F=t, \mu_1<0, \mu_2 > 0$ (the \cite{PhysRevLett.124.196603} construction). Indices $1,2$ denote the two nodes.
The Matsubara Green's function for node 1 is
\begin{equation}
\label{G_1}
\begin{split}
    G_1(i\epsilon_n,\mathbf{k})&=\frac{1}{i\epsilon
    _n-v_F\mathbf{k}\cdot \boldsymbol{\sigma}+\mu_1}\\
    &=\frac{P_+(\mathbf{k})}{i\epsilon_n-v_Fk+\mu_1}+\frac{P_-(\mathbf{k})}{i\epsilon_n+v_Fk+\mu_1},
\end{split}
\end{equation}
where $P_\pm(\mathbf{k})\equiv (I\pm \hat{\mathbf{k}}\cdot \boldsymbol{\sigma})/2$ with $\hat{\mathbf{k}}=\mathbf{k}/k$ projects on the eigenstate of the Dirac operator with helicity $\pm$. Note also that $P^2_\pm(\mathbf{k})=P_\pm(\mathbf{k})$. Similarly, for node 2, we find the following
\begin{equation}
\label{G_2}
G_2(i\epsilon_n,\mathbf{k})=\frac{P_+(\mathbf{k})}{i\epsilon_n+v_Fk+\mu_2}+\frac{P_-(\mathbf{k})}{i\epsilon_n-v_Fk+\mu_2}.
\end{equation}
Note that due to the minus sign in the Hamiltonian $(-v_F\mathbf{k}\cdot \boldsymbol{\sigma}-\mu_2)$, we have a reverse sign for the upper and lower bands for node 2.
This effective low-energy theory for the Weyl semi-metal is obtained by setting a momentum cutoff $|\mathbf{k}-\mathbf{k}_a|<\Lambda$ around each Weyl node $a$.
Adding the onsite random scalar potential
\begin{equation}
    H=H_0+H_\mathrm{dis}=\sum_\mathbf{k}H_0(\mathbf{k})+\sum_\mathbf{r} \Psi^\dagger(\mathbf{r}) V(\mathbf{r}) \Psi(\mathbf{r})
\end{equation}
where $V(\mathbf{r})$ is drawn from a Gaussian distribution $\mathcal{N}(0,W^2)$ with zero average and variance $W^2$:
\begin{equation}
    \overline{V(\mathbf{r})}=0,\quad \overline{V(\mathbf{r})V(\mathbf{r}')}=W^2\delta(\mathbf{r}-\mathbf{r}'). 
\end{equation}

\begin{figure}[t!]
\begin{center}
\includegraphics[width=0.6\linewidth]{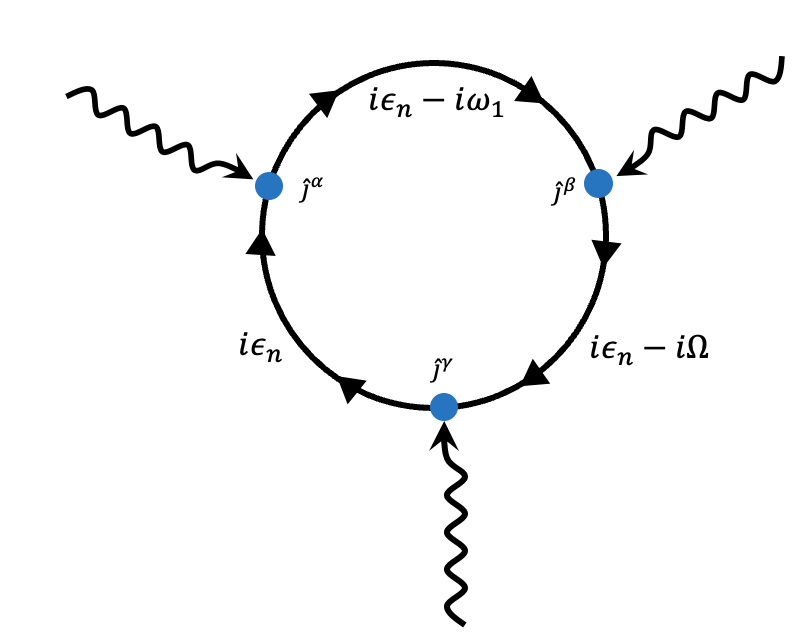}
\caption{The Feynman diagrams for the bare second order conductivity. Lines with arrows represent Fermion fields and Green's functions with certain frequency. Curly curves represent external fields.}
\label{vertex}
\end{center}
\end{figure}

\begin{figure}[t!]
\begin{center}
\setlabel{pos=nw,fontsize=\large,labelbox=false}
\xincludegraphics[scale=0.45,label=a]{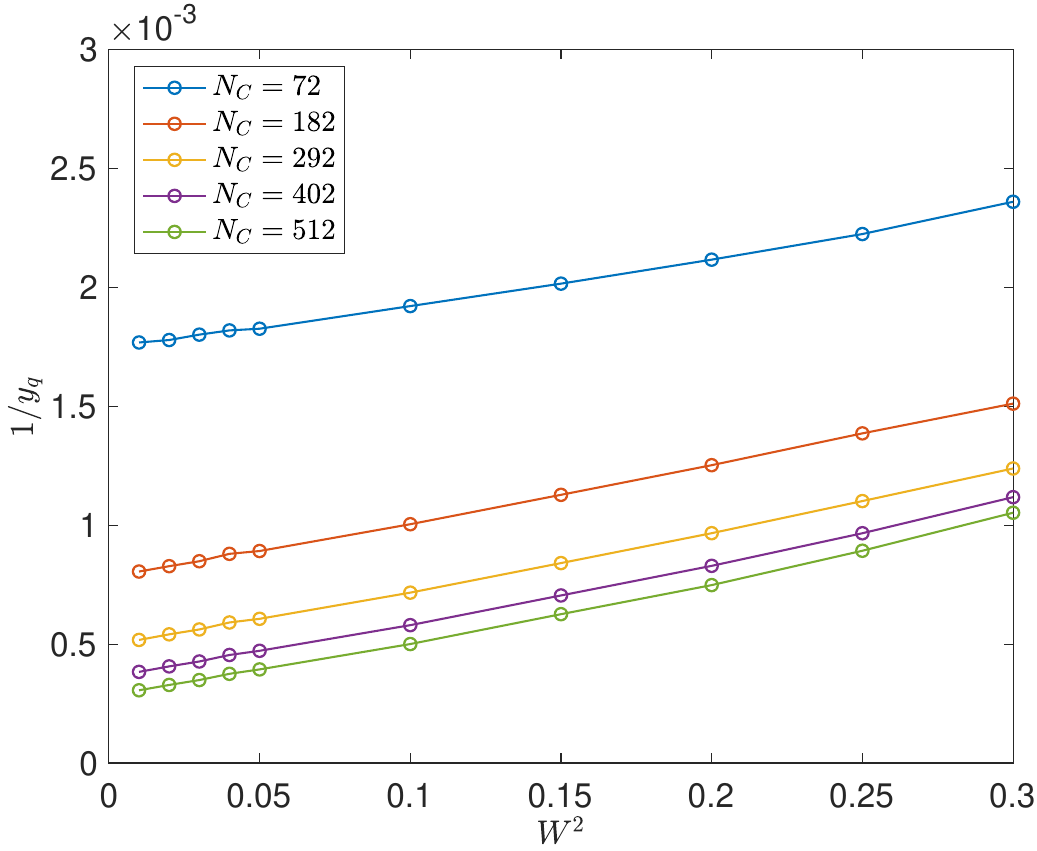}
\xincludegraphics[scale=0.45,label=b]{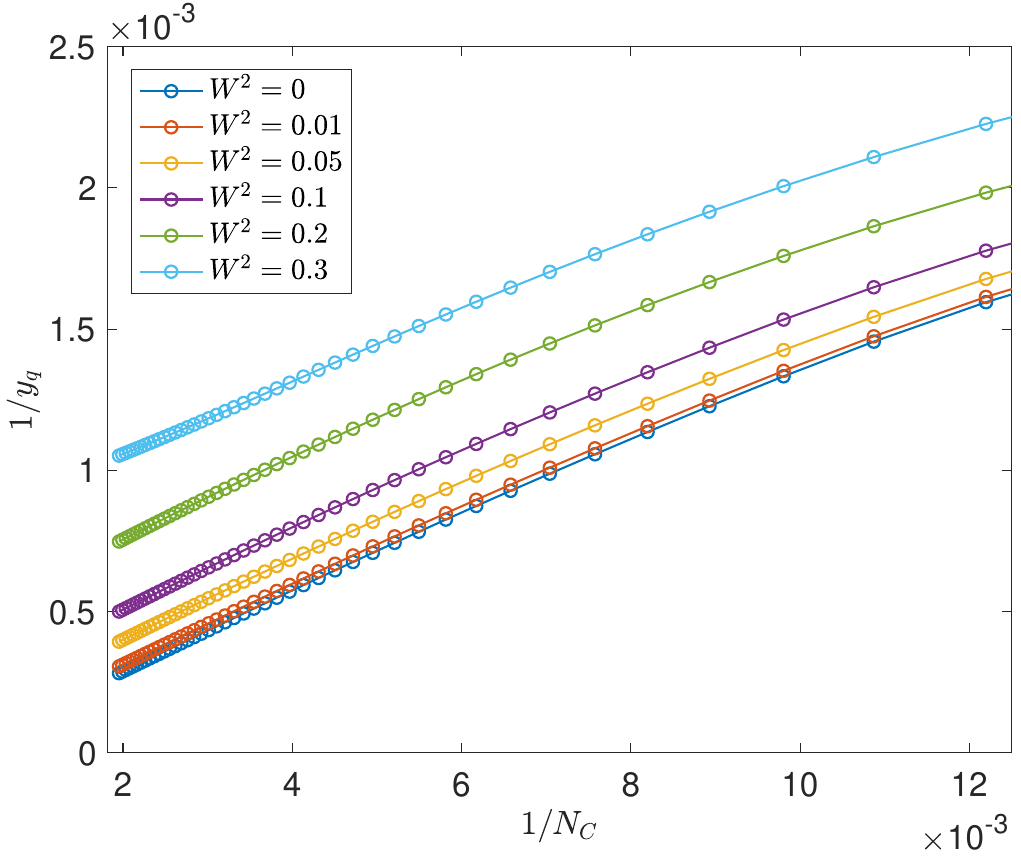}
\caption{(a) The scaling of the plateau value $1/y_q$ versus disorder strength $W^2$, where the $W^2$ is the variance of the Gaussian disorder. We take $\omega_1=\gamma=0.8t$ to have relatively large plateau. $y_q$ value is taken at approximately the maximal value of the plateau at around $E_F=-0.1$. (b) The scaling of the plateau value $1/y_q$ versus expansion order $1/N_C$ for vairous disorder strengths (same data from (a)). The disordered results are averaged over 50 realizations. }
\label{fig:disorderscale}
\end{center}
\end{figure}

Our aim is to compute the second order in $W$ correction to the CPGE which is obtained from the second-order conductivity
\begin{equation}
    \begin{split}
        &j^\gamma(\Omega)=\frac{\chi^{\alpha\beta\gamma}(\omega_1,\omega_2)+\chi^{\beta\alpha\gamma}(\omega_2,\omega_1)}{\omega_1\omega_2}E^\alpha(\omega_1) E^\beta(\omega_2),\\
        &\chi^{\alpha\beta\gamma}(i\omega_1,i\omega_2)=T\sum_{\varepsilon_n}\int \frac{\mathrm{d}^3k}{(2\pi)^3} \mathrm{Tr}[\hat{j}^\alpha G(i\varepsilon_n-i\omega_1,\mathbf{k}) \\
        &\hat{j}^\beta G(i\varepsilon_n-i\Omega,\mathbf{k})\hat{j}^\gamma G(i\varepsilon_n,\mathbf{k})],
    \end{split}
\end{equation}
where $\Omega=\omega_1+\omega_2\to 0$ and the diagram for $\chi$ is shown in Fig. \ref{vertex}(b). Note that in the basis $(\psi_1,\psi_2)^T$ the bare current operator for Weyl fermions is given by 
\begin{equation}
    \hat{j}^\alpha =e\frac{\delta \hat{H}_0(\mathbf{k})}{\delta k^\alpha} =ev_F \begin{pmatrix}
        \sigma^\alpha & 0\\
        0 & -\sigma^\alpha
    \end{pmatrix}.
\end{equation}
Before moving on, we observe that the numerical results from the KPM indicate that the bare plateau value $y_q$ of the function
\begin{equation}
    y\equiv \mathrm{Im}\bigg[ \frac{\chi^{\alpha\beta\gamma}(\omega_1,\omega_2)+\chi^{\beta\alpha\gamma}(\omega_2,\omega_1)}{\omega_1\omega_2} \bigg],
\end{equation}
scales linearly with the expansion order $N_C$ as shown in Fig. \ref{fig:NCscale}. For a fixed $N_C$, the results $y_q$ are stabilized for large system sizes $L$. As we already take the bare value $\Omega=0,$ or $\omega_2=-\omega_1$, the analytical divergence of the quantization $\beta_0=y_q/\Omega$ is now governed by the finite broadening $O(1/N_C)$ in the Green's function (finite life time $O(N_C)$) introduced via the KPM. It is known that for the disordered electron gas (metal), the disorder could also induce a finite lifetime via self-energy correction, which can also change the value of $y_q$. Numerically in Fig. \ref{fig:disorderscale}, we observed that there is a clear linear scaling of $1/y_q$ versus the variance of disorder $W^2$. In order to understand this results, we study the first-order perturbation theory with disorder upon the $\chi^{\alpha\beta\gamma}$ below.

\begin{figure}[t!]
\begin{center}
\includegraphics[width = 0.35\textwidth]{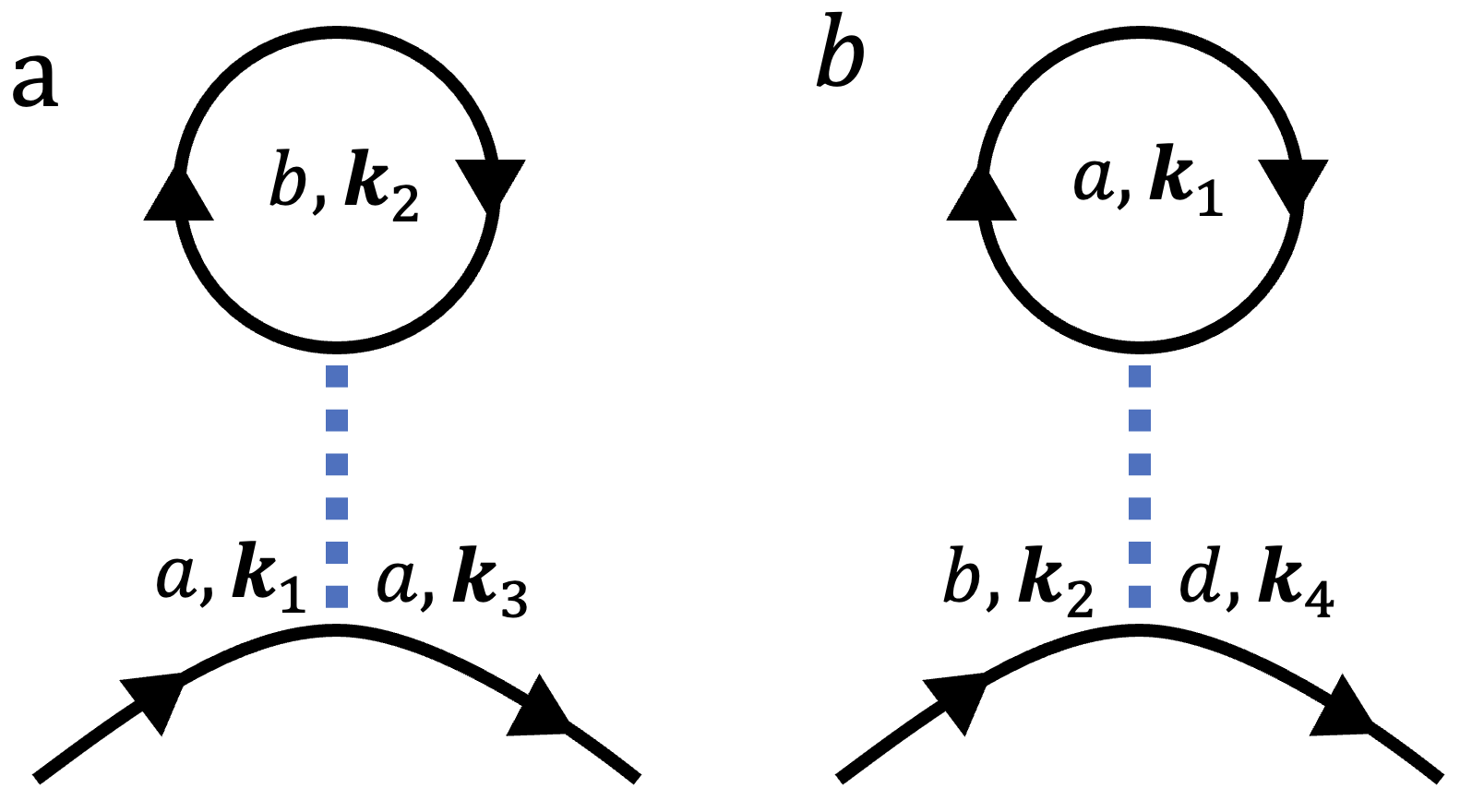}
\includegraphics[width = 0.45\textwidth]{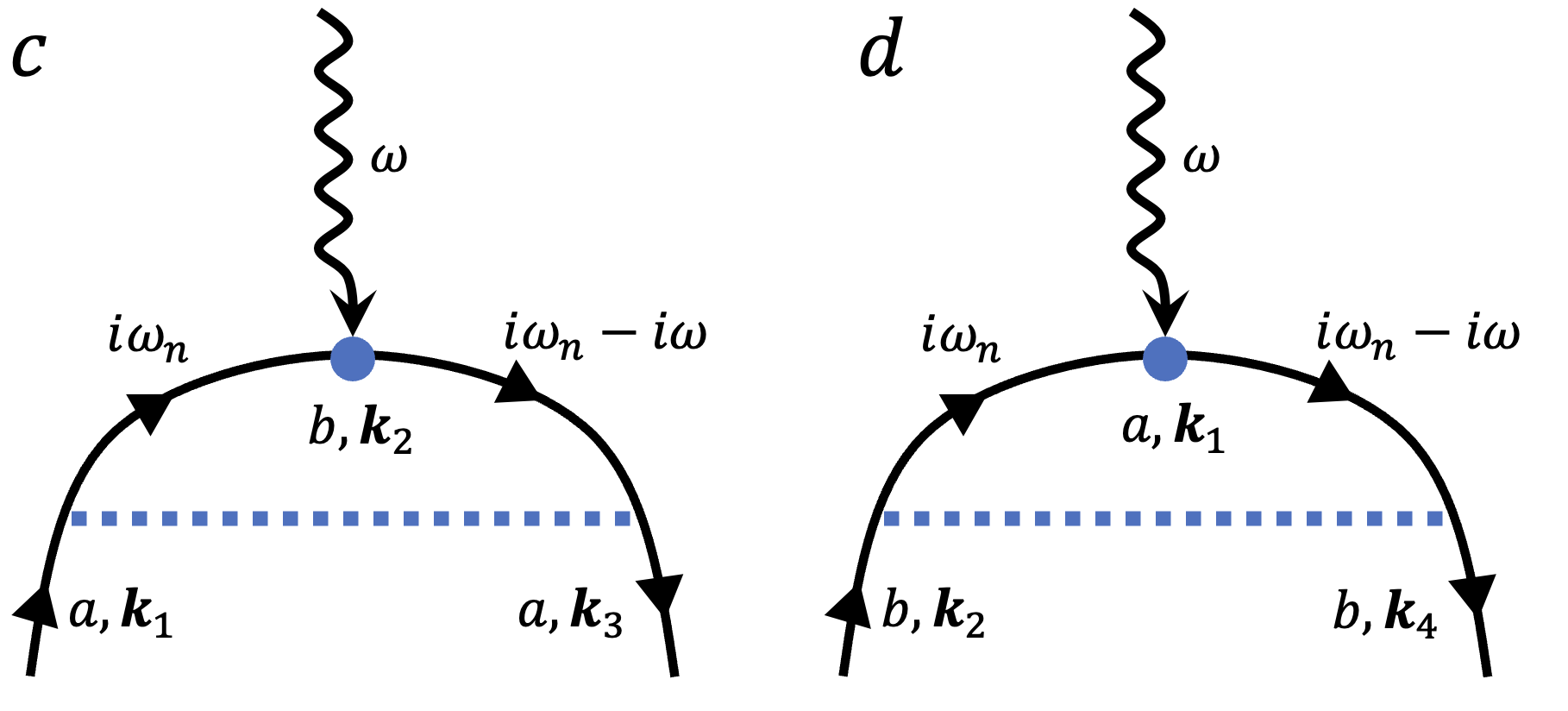}
\caption{The Feynman diagrams for the first-order correction to the $\chi^{\alpha\beta\gamma}$. (a-b) second-order corrections (bubble diagrams) to the three single-particle Green's functions. (c-d) Vertex corrections to the current.  }
\label{1storder}
\end{center}
\end{figure}

The lowest order correction to $\chi^{\alpha\beta\gamma}$ gives 12 contributions (diagrams), where 6 of them are corrections to the single Green's function with three frequencies as shown in Fig. \ref{1storder} (a-d) and the rest are vertex corrections to the current as shown in Fig. \ref{1storder} (e-h). We now compute the self-energy correction given by disorder-scattering processes. Within the lowest-order Born approximation, we find:
\begin{equation}
\label{Dyson_equation}
\Sigma(i\epsilon) =  \frac{W^2}{N}\sum_{\mathbf k} G(\mathbf k,i\epsilon)=\frac{W^2}{N}\sum^\Lambda_{\mathbf k}\sum_{l} G_l(\mathbf k,i\epsilon),
\end{equation}
where we consider the integral over momentum is performed in a circle of radius $\Lambda$ around the two low energy Weyl nodes. 
We solve the Dyson equation~\eqref{Dyson_equation} to the lowest order in $W^2$, which consists of replacing $G$ in Eq.~\eqref{Dyson_equation} with the bare Green's function given in Eqs.~\eqref{G_1} and~\eqref{G_2}. To make further progress, we compute the integral: 
\begin{equation}
\begin{split}
    F_l(i\epsilon)&=\frac{1}{N}\sum_{\mathbf{k}} G_{l}(\mathbf{k},i\epsilon)\\
    &=\frac{1}{2}\sum_{\eta=\pm }\int \frac{d^3\mathbf{k}}{\Omega_{\rm BZ}} \left[\mathbf{1}+\eta\frac{\mathbf{k}}{|\mathbf{k}|}\cdot{\pmb{ \sigma}}\right]\frac{1}{i\epsilon-\eta v_{Fl} |\mathbf{k}| +\mu_l}, 
\end{split}
\end{equation}
where $v_{F1}=-v_{F2}=v_F$ and  $T_l(i\epsilon)$ is a $2\times2$ matrix in the spin degrees of freedom. Performing the angular integral, we find: 
\begin{equation}
    F_l(i\epsilon)=\sum_{\eta=\pm}\frac{2\pi}{\Omega_{\rm BZ}}\int^\Lambda_0 dk \frac{k^2}{i\epsilon-\eta v_{F l}k +\mu_l}.
\end{equation}
Notice that the UV cut-off must be introduced since the integrand is not bounded and diverges for large $k$. The integral over $k$ can be performed analytically. It is convenient to perform the integral employing the approximation: 
\begin{equation}
    \frac{1}{i\epsilon - \eta v_{Fl}k+\mu_l} = \frac{1}{\mu_l -\eta v_{Fl} k}-i\pi \text{sign}\epsilon \delta(\eta v_{Fl}k-\mu_l).
\end{equation}
Employing the latter expression, we get:
\begin{equation}
    F_l(i\epsilon) =\frac{4\mu_l\pi}{\Omega_{\rm BZ}}P\int^\Lambda_0\frac{k^2}{\mu^2_l-v^2_{Fl}k^2}-i\pi\frac{2\pi}{\Omega_{\rm BZ}}\text{sign}(\epsilon)\frac{\mu^2_l}{v^3_{Fl}}.
\end{equation}
In the limit of small chemical potential, we are left with a simple contribution: 
\begin{equation}
    F_l(i\epsilon)=\rho(\mu_l)\left[-\frac{\Lambda v_{Fl} }{\mu_l}+\text{Atanh}\frac{v_{Fl}\Lambda}{\mu_l}\right]-\frac{i\pi}{2}\text{sign}(\epsilon)\rho(\mu_l),
\end{equation}
which gives the self-energy: 
\begin{equation}
\label{Dyson_equation_1}
\Sigma(i\epsilon) =  W^2\sum_{a=1}^2 F_a(i\epsilon).
\end{equation}
In the previous expression we have introduced the self-energy 
\begin{equation}
    \rho(\omega) = \frac{3\epsilon^2}{2(v_{F}\Lambda)^2},
\end{equation}
with $\epsilon\in[-v_F\Lambda,v_F\Lambda]$.

The vertex corrections can be similarly carried out. For a single current $\hat{j}^\alpha$, the vertex correction for each node is
\begin{equation}
\begin{split}
     [\delta j^\alpha]_1&=\frac{W^2}{2N} \sum_\mathbf{k}G^{(1)}(i\epsilon_n,\mathbf{k})ev_F\sigma^\alpha G^{(1)}(i\epsilon_n-i\omega_1,\mathbf{k}).
\end{split}
\end{equation}
\begin{equation}
\begin{split}
     [\delta j^\alpha]_2&=\frac{W^2}{2N} \sum_\mathbf{k}G^{(2)}(i\epsilon_n,\mathbf{k})(-1)ev_F\sigma^\alpha G^{(2)}(i\epsilon_n-i\omega_1,\mathbf{k}).
\end{split}
\end{equation}
We notice that the cross term vanishes to $W^2$ order in the perturbation theory. With the prototype integral $\int \frac{x^2dx}{(a+bx)(c+dx)}=\frac{x}{bd}+\frac{a^2\ln(a+bx)}{b^2(bc-ad)}-\frac{c^2\ln(c+dx)}{d^2(bc-ad)}$, consider the 4 integrals
\begin{widetext}
\begin{equation}
\begin{split}
    &\int_0^\Lambda \frac{4\pi k^2 dk}{(2\pi)^3} \frac{1}{i\epsilon_n-v_Fk+\mu_1}\frac{1}{i\epsilon_n-v_Fk+\mu_1-\omega_1}\\
    &=\frac{1}{2\pi^2}\bigg[\frac{\Lambda}{v_F^2}+\frac{(i\epsilon_n+\mu_1)^2}{v_F^3\omega_1}\ln\frac{i\epsilon_n+\mu_1-v_F\Lambda}{i\epsilon_n+\mu_1}-\frac{(i\epsilon_n+\mu_1-\omega_1)^2}{v_F^3\omega_1}\ln\frac{i\epsilon_n+\mu_1-\omega_1-v_F\Lambda}{i\epsilon_n+\mu_1-\omega_1}\bigg]\\
    &\int_0^\Lambda \frac{4\pi k^2 dk}{(2\pi)^3} \frac{1}{i\epsilon_n-v_Fk+\mu_1}\frac{1}{i\epsilon_n+v_Fk+\mu_1-\omega_1}\\
    &=\frac{1}{2\pi^2}\bigg[-\frac{\Lambda}{v_F^2}-\frac{(i\epsilon_n+\mu_1)^2}{v_F^3(2i\epsilon_n+2\mu_1-\omega_1)}\ln\frac{i\epsilon_n+\mu_1-v_F\Lambda}{i\epsilon_n+\mu_1}+\frac{(i\epsilon_n+\mu_1-\omega_1)^2}{v_F^3(2i\epsilon_n+2\mu_1-\omega_1)}\ln\frac{i\epsilon_n+\mu_1-\omega_1+v_F\Lambda}{i\epsilon_n+\mu_1-\omega_1}\bigg]\\
    &\int_0^\Lambda \frac{4\pi k^2 dk}{(2\pi)^3} \frac{1}{i\epsilon_n+v_Fk+\mu_1}\frac{1}{i\epsilon_n-v_Fk+\mu_1-\omega_1}\\
    &=\frac{1}{2\pi^2}\bigg[-\frac{\Lambda}{v_F^2}+\frac{(i\epsilon_n+\mu_1)^2}{v_F^3(2i\epsilon_n+2\mu_1-\omega_1)}\ln\frac{i\epsilon_n+\mu_1+v_F\Lambda}{i\epsilon_n+\mu_1}-\frac{(i\epsilon_n+\mu_1-\omega_1)^2}{v_F^3(2i\epsilon_n+2\mu_1-\omega_1)}\ln\frac{i\epsilon_n+\mu_1-\omega_1-v_F\Lambda}{i\epsilon_n+\mu_1-\omega_1}\bigg]\\
    &\int_0^\Lambda \frac{4\pi k^2 dk}{(2\pi)^3} \frac{1}{i\epsilon_n+v_Fk+\mu_1}\frac{1}{i\epsilon_n+v_Fk+\mu_1-\omega_1}\\
    &=\frac{1}{2\pi^2}\bigg[\frac{\Lambda}{v_F^2}-\frac{(i\epsilon_n+\mu_1)^2}{v_F^3 \omega_1}\ln\frac{i\epsilon_n+\mu_1+v_F\Lambda}{i\epsilon_n+\mu_1}+\frac{(i\epsilon_n+\mu_1-\omega_1)^2}{v_F^3\omega_1}\ln\frac{i\epsilon_n+\mu_1-\omega_1+v_F\Lambda}{i\epsilon_n+\mu_1-\omega_1}\bigg]
\end{split}
\end{equation}
\end{widetext}
Summing over these integrals gives
\begin{widetext}
\begin{equation}
\begin{split}
     &[\delta j^\alpha]_1=\frac{ev_F\sigma^\alpha W^2}{2\pi^2} \bigg[-\frac{2(i\epsilon_n+\mu_1)^2(i\epsilon_n+\mu_1-\omega_1)}{v_F^3\omega_1(2i\epsilon_n+2\mu_1-\omega_1)}\ln\frac{i\epsilon_n+\mu_1+v_F\Lambda}{i\epsilon_n+\mu_1-v_F\Lambda}\\
     &+\frac{2(i\epsilon_n+\mu_1-\omega_1)^2(i\epsilon_n+\mu_1)}{v_F^3\omega_1(2i\epsilon_n+2\mu_1-\omega_1)}\ln\frac{i\epsilon_n+\mu_1-\omega_1+v_F\Lambda}{i\epsilon_n+\mu_1-\omega_1-v_F\Lambda}\bigg]
\end{split}
\end{equation}
\end{widetext}
Sending $\mu_1\to \mu_1-\omega_1, \omega_1\to -\omega_1$, we have
\begin{widetext}
\begin{equation}
\begin{split}
     &[\delta j^\beta]_1=\frac{ev_F\sigma^\beta W^2}{2\pi^2} \bigg[-\frac{2(i\epsilon_n+\mu_1-\omega_1)^2(i\epsilon_n+\mu_1)}{-v_F^3\omega_1(2i\epsilon_n+2\mu_1-\omega_1)}\ln\frac{i\epsilon_n+\mu_1-\omega_1+v_F\Lambda}{i\epsilon_n+\mu_1-\omega_1-v_F\Lambda}\\
     &+\frac{2(i\epsilon_n+\mu_1)^2(i\epsilon_n+\mu_1-\omega_1)}{-v_F^3\omega_1(2i\epsilon_n+2\mu_1-\omega_1)}\ln\frac{i\epsilon_n+\mu_1+v_F\Lambda}{i\epsilon_n+\mu_1-v_F\Lambda}\bigg]
\end{split}
\end{equation}
\end{widetext}
$\omega_1\to 0$ gives $[j^\gamma]^{(f)}=0$. Thus, combine the contribution from each current, the correction to the CPGE response due to node $1$
\begin{widetext}
\begin{equation}
\begin{split}
     &[\delta\beta(\omega_1)]_1=\frac{\beta_0 W^2}{2\pi^2} \bigg[\frac{4(i\epsilon_n+\mu_1-\omega_1)^2(i\epsilon_n+\mu_1)}{v_F^3\omega_1(2i\epsilon_n+2\mu_1-\omega_1)}\ln\frac{i\epsilon_n+\mu_1-\omega_1+v_F\Lambda}{i\epsilon_n+\mu_1-\omega_1-v_F\Lambda}\\
     &-\frac{4(i\epsilon_n+\mu_1)^2(i\epsilon_n+\mu_1-\omega_1)}{v_F^3\omega_1(2i\epsilon_n+2\mu_1-\omega_1)}\ln\frac{i\epsilon_n+\mu_1+v_F\Lambda}{i\epsilon_n+\mu_1-v_F\Lambda}\bigg]
\end{split}
\end{equation}
\end{widetext}
And similarly for node $2$, the opposite monopole charge has opposite correction.
\begin{widetext}
\begin{equation}
\begin{split}
     &[\delta \beta(\omega_1)]_2=\frac{-\beta_0 W^2}{2\pi^2} \bigg[\frac{4(i\epsilon_n+\mu_2-\omega_1)^2(i\epsilon_n+\mu_2)}{v_F^3\omega_1(2i\epsilon_n+2\mu_2-\omega_1)}\ln\frac{i\epsilon_n+\mu_2-\omega_1+v_F\Lambda}{i\epsilon_n+\mu_2-\omega_1-v_F\Lambda}\\
     &-\frac{4(i\epsilon_n+\mu_2)^2(i\epsilon_n+\mu_2-\omega_1)}{v_F^3\omega_1(2i\epsilon_n+2\mu_2-\omega_1)}\ln\frac{i\epsilon_n+\mu_2+v_F\Lambda}{i\epsilon_n+\mu_2-v_F\Lambda}\bigg]
\end{split}
\end{equation}
\end{widetext}
In the limit of $i\epsilon_n\to 0,v_F\Lambda \gg \mu_1,\omega_1$,
\begin{widetext}
\begin{equation}
    \begin{split}
        &[\delta\beta(\omega_1)]_1\approx \beta_0\frac{2W^2}{\pi^2}\bigg[\frac{(\mu_1-\omega_1)^2\mu_1}{v_F^3\omega_1(2\mu_1-\omega_1)} (2\frac{\mu_1-\omega_1}{v_F\Lambda})-\frac{\mu_1^2(\mu_1-\omega_1)}{v_F^3\omega_1(2\mu_1-\omega_1)}(2\frac{\mu_1}{v_F\Lambda})\bigg]=-\beta_0\frac{4W^2}{\pi^2}\frac{\mu_1(\mu_1-\omega_1)}{v_F^4\Lambda}\\
        &\delta \beta(\omega_1) = -\beta_0\frac{4W^2}{\pi^2v_F^4\Lambda}[\mu_1(\mu_1-\omega_1)-\mu_2(\mu_2-\omega_1)]=-\beta_0\frac{4W^2}{\pi^2v_F^4\Lambda}[\mu_1^2-\mu_2^2+\omega_1(\mu_2-\mu_1)]
    \end{split}
\end{equation}
\end{widetext}

\clearpage
\section{Numerical Disorder dependence of the CPGE} \label{App:C}

\begin{figure*}[t!]
\begin{center}
\setlabel{pos=nw,fontsize=\large,labelbox=false}
\xincludegraphics[scale=0.33,label=a]{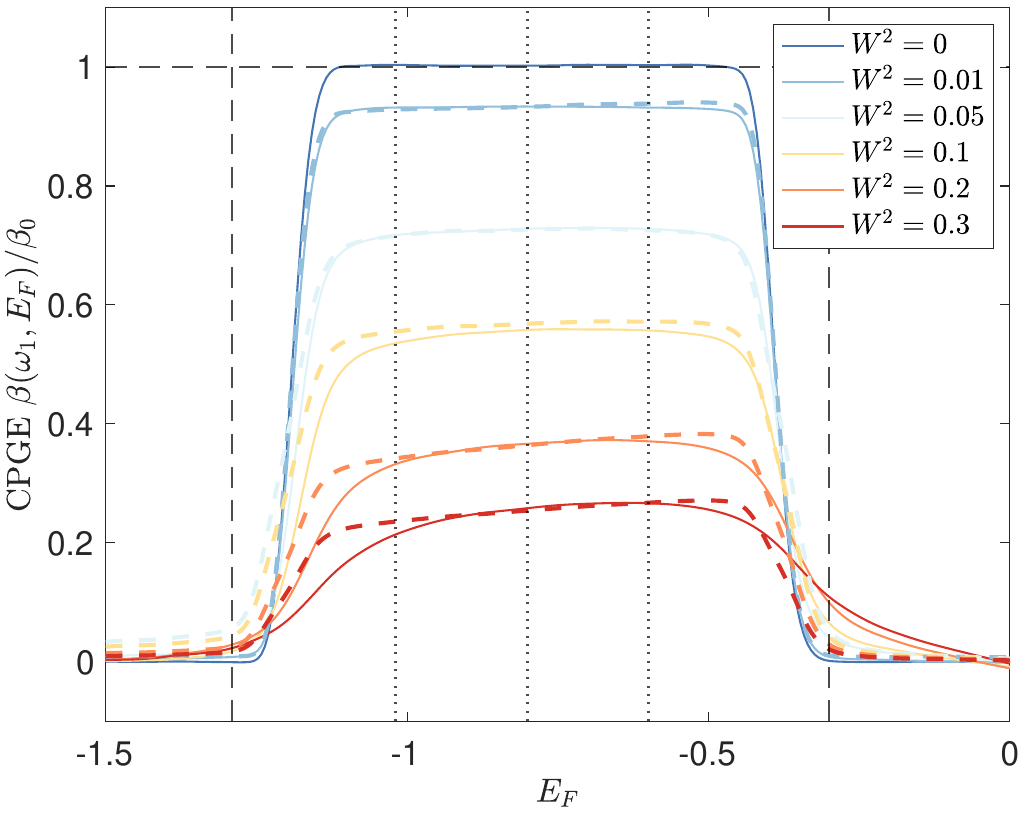}
\xincludegraphics[scale=0.33,label=b]{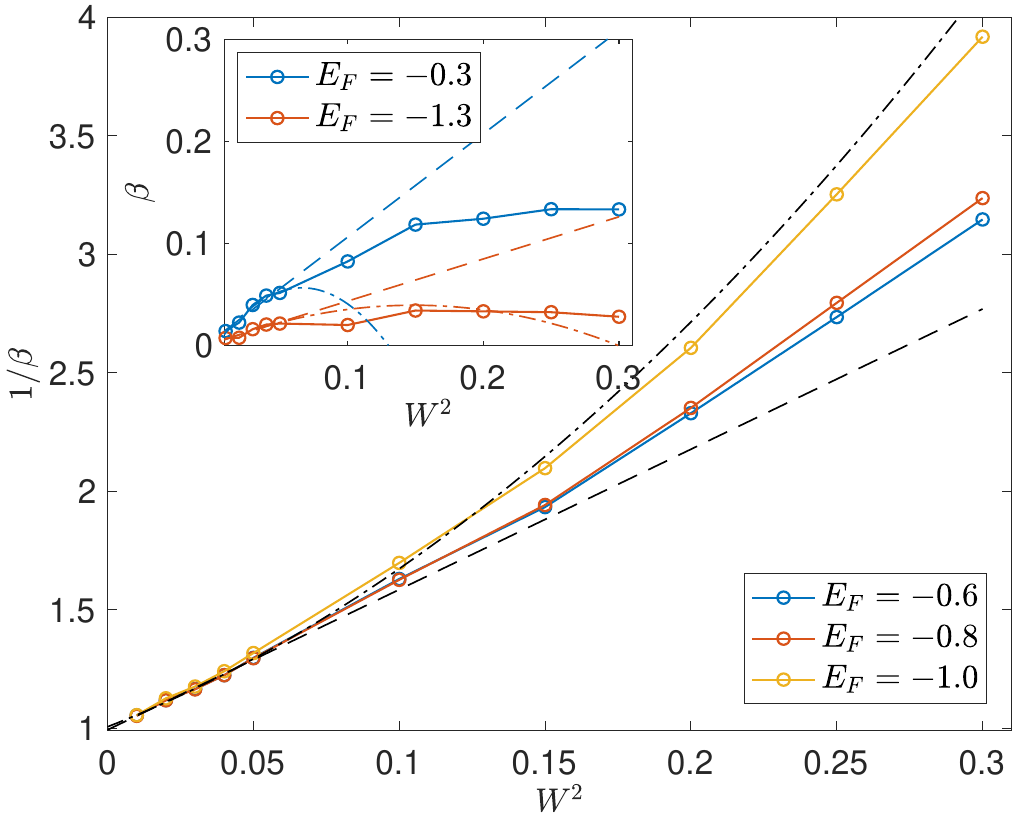}
\xincludegraphics[scale=0.33,label=c]{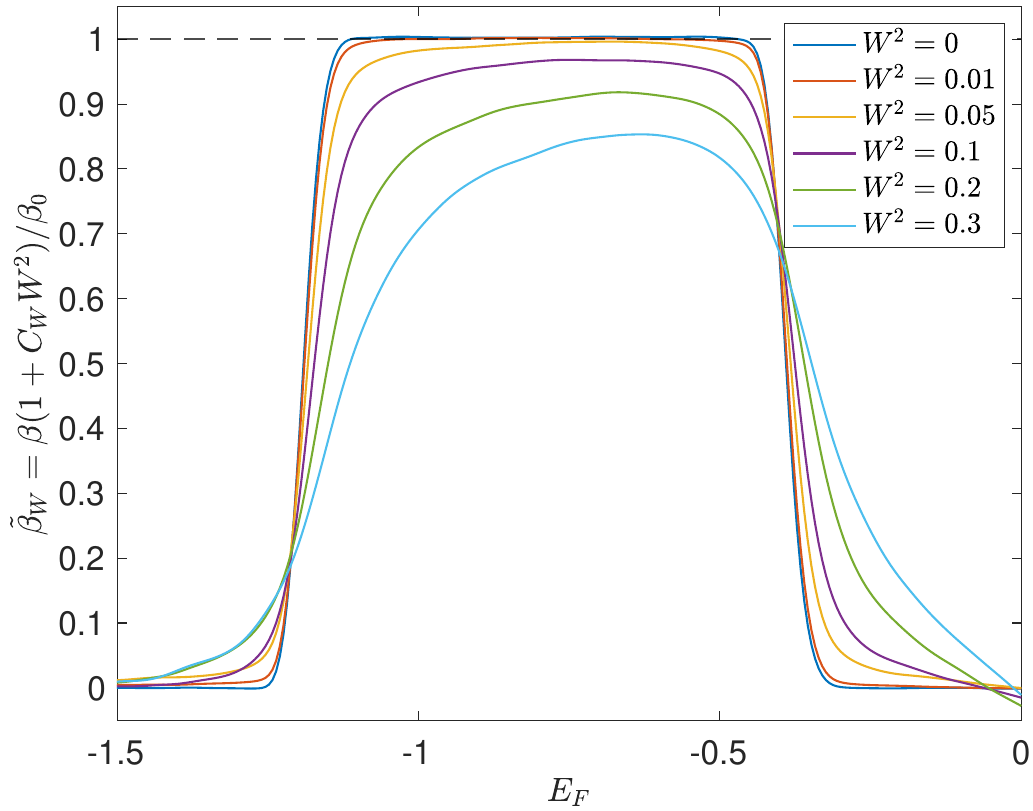}
\caption{ (a) The CPGE quantization in the clean limit and its suppression with increasing disorder strength with $L=100, N_C=2^9$. In KPM, the half-bandwidth $D=6t$ to fit with various disorder strength.  Each disorder result $W$ is averaged over $50$ disorder realizations and twisted boundary conditions. The normalization factor is the same as in the clean case $C_\mathrm{KPM}=31.74\beta_0$. Dashed curves mark perturbative corrections with quasiparticle lifetime, frequency shift, and vertex correction. Vertical dotted lines mark locations $E_F=[-1.0,-0.8,-0.6]t$, while vertical dashed lines mark $E_F=[-1.3,0.3]t$. 
(b) The normalized CPGE response $1/\beta$ as a function of disorder strength $W^2$ at various Fermi levels with $N_C=512$ from marked $E_F$. The dashed curve shows a linear fit to the first $5$ small $W$ of the middle curve, where $\beta_0/\beta \approx 0.999+C_W W^2,C_W=7.39$. The dash-dotted curves are the second-order polynomial fit with first $5$ points.
Inset: The disorder broadening $\beta$ at two different $E_F$ (plateau edges) from (a), which have vanished responses from the clean systems. The dashed curves are linear fits from the first $5$ data points and dash-dotted curves are second-order polynomial fits. (c) The scaled CPGE response with response to disorder, $\tilde{\beta}_W\equiv \beta(1+C_WW^2)/\beta_0$, from (a).
}
\label{fig:disorder}
\end{center}
\end{figure*}

\begin{figure*}[b!]
\centering
\setlabel{pos=nw,fontsize=\large,labelbox=false}
\xincludegraphics[scale=0.4,label=a]{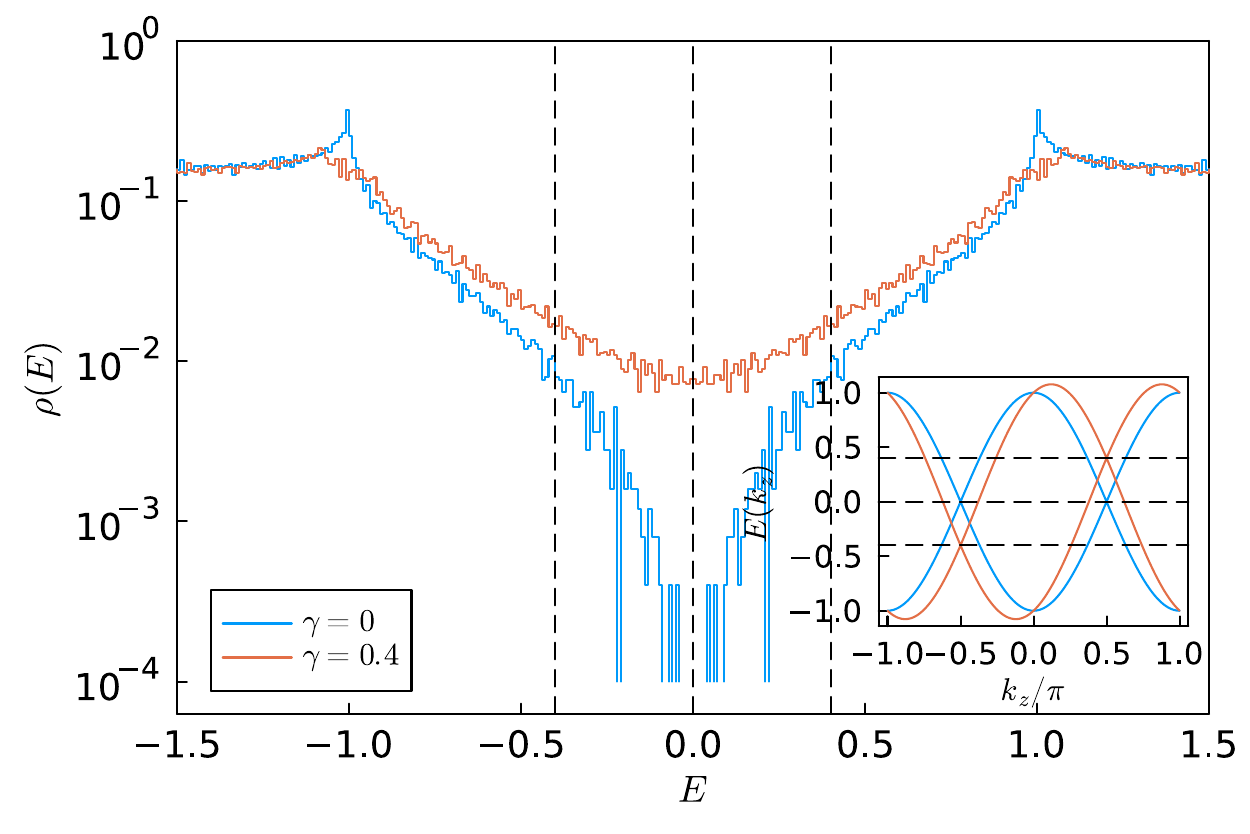}
\xincludegraphics[scale=0.4,label=b]{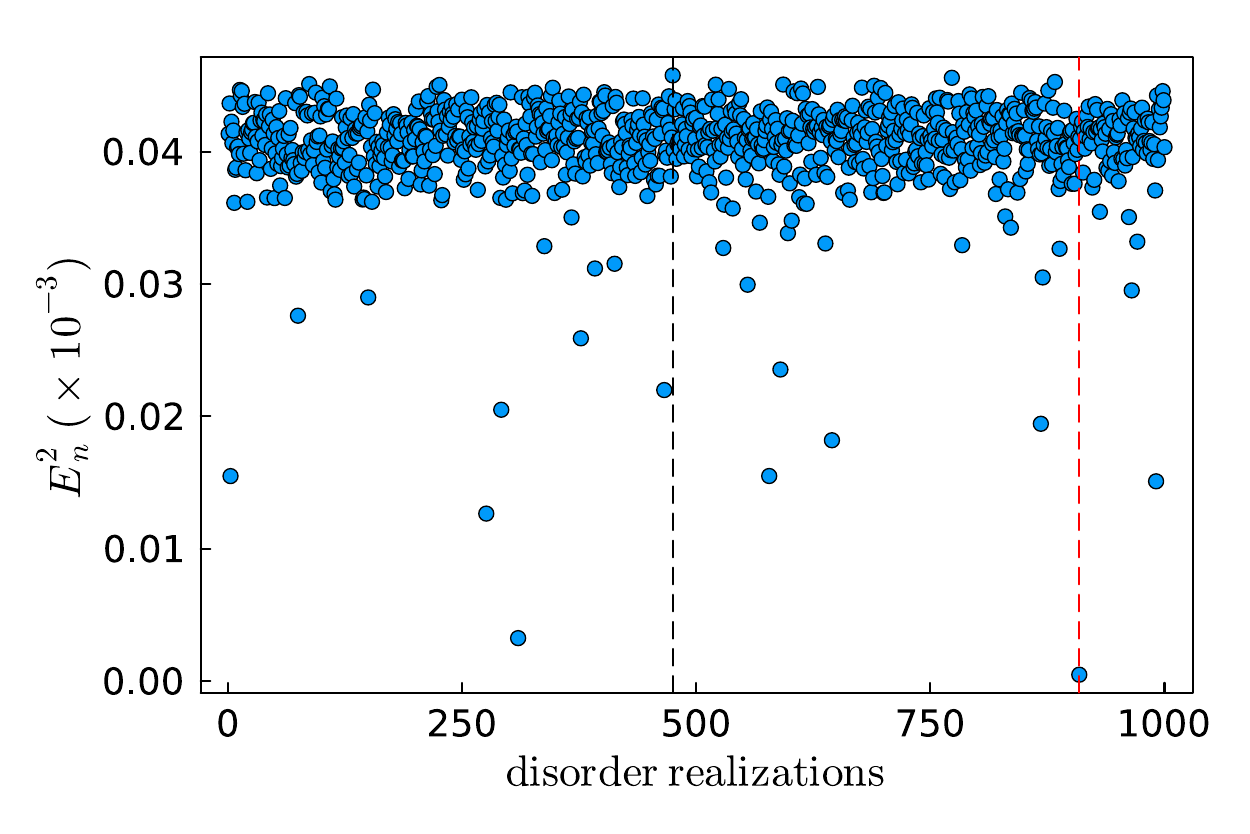}
\xincludegraphics[scale=0.4,label=c]{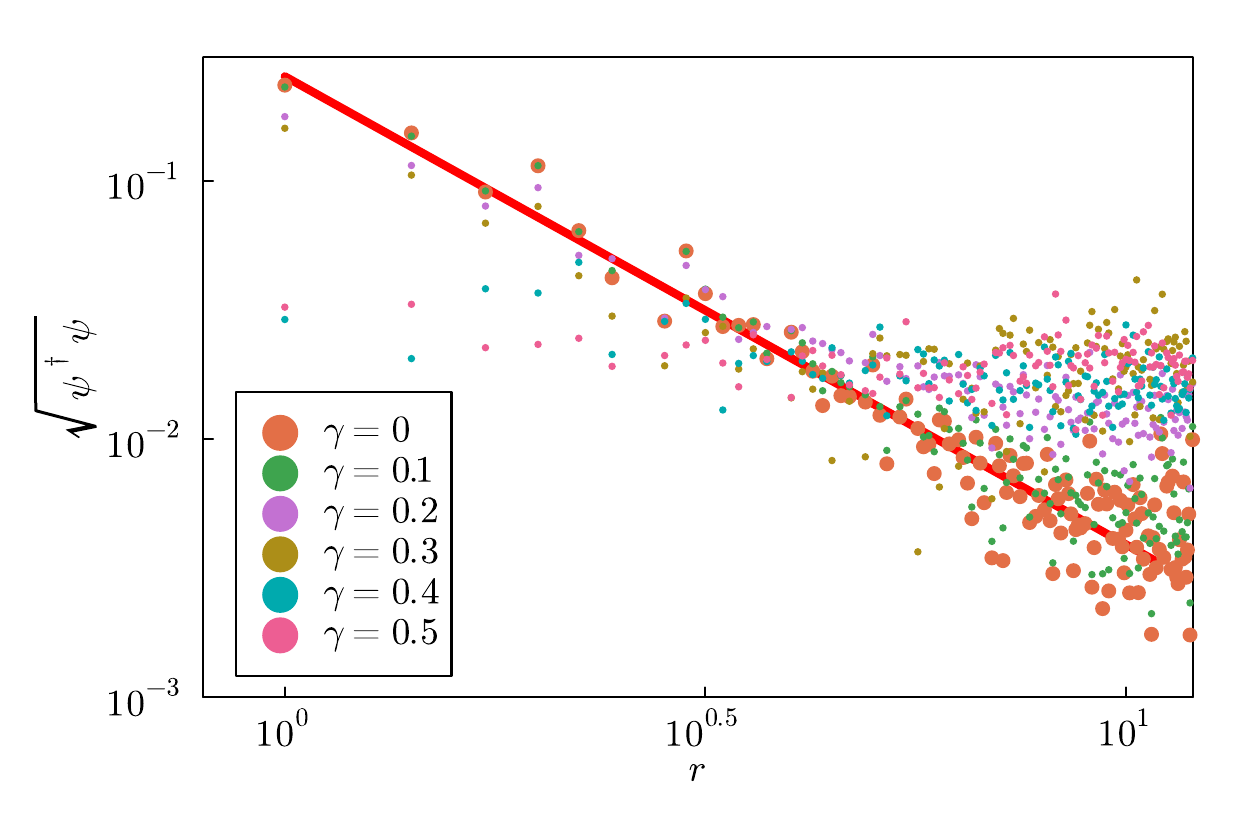}
\xincludegraphics[scale=0.4,label=d]{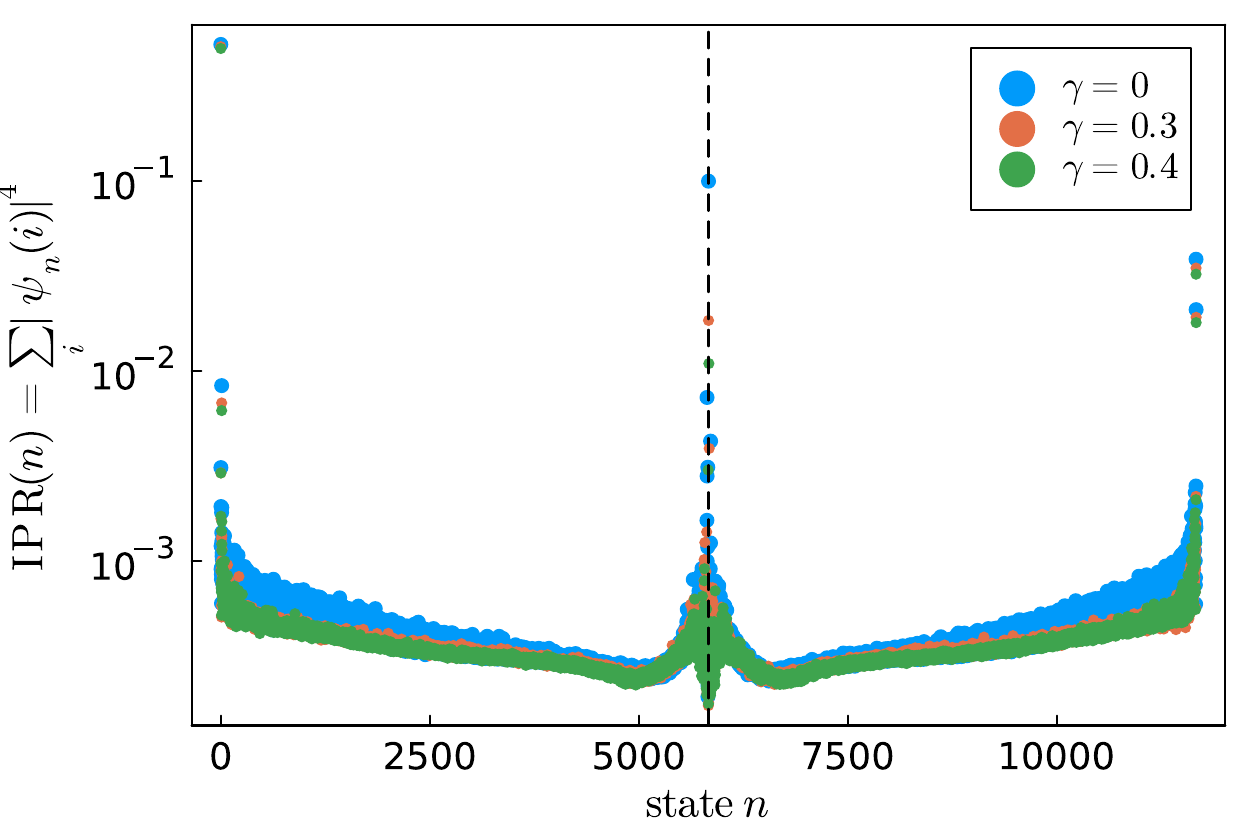}
\caption{Rare states from increasing $\gamma$ for lattice systems $L=18$. (a) The density of states with and without finite $\gamma$. Inset: the corresponding dispersion for the two $\gamma$ cases. (b) The squared eigenvalues across the disorder realizations. Vertical dashed lines mark the rare state sample (red) and the perturbative sample (black). (c) The power-law decay of the wavefunction as a function of the real space distance $r$ from the maximum values of the eigenstates with various $\gamma$. The red line marks the power law fitting. (d) The IPR of eigenstates for various $\gamma$.}
\label{fig:rare}
\end{figure*}

The disorder dependence of the CPGE behaves differently at various Fermi energies $E_F$ as shown in Fig. \ref{fig:disorder}. At the edges of the quantization plateau, $E_F=-1.3,0.3$, the CPGE response linearly correlates with disorder
\begin{equation}
    \beta \approx \beta^{(0)}+\frac{\beta^{(2)}}{2!} W^2+\cdots,
\end{equation}
where $\beta^{(n)}$ is the $n$th derivative of $\beta$ with respective to $W$ and $\beta^{(2)}>0$. In contrast, at the quantized plateau, $E_F=-1.0,-0.8,-0.6$, the CPGE decreases linearly, $\beta^{(2)}<0$. From the perspective of quasi-particle lifetime, the plateau value of the CPGE scales inversely with disorder $W^2$
\begin{equation}
\begin{split}
     \frac{\beta}{\beta_0} &=\frac{1}{1+C W^2} \to \frac{1}{\beta} =\frac{1}{\beta_0}+\frac{C}{\beta_0}W^2 \\
    &\approx \frac{1}{\beta^{(0)}} +\frac{1}{2!}\bigg[\frac{2\beta^{(1)}}{(\beta^{(0)})^3}-\frac{\beta^{(2)}}{(\beta^{(0)})^2}\bigg] W^2+\cdots,
\end{split}
\end{equation}
where the inverse dependence is confirmed in Fig. \ref{fig:disorderscale} $\mathbf{b}$ at small $W^2$ limit. When we rescale the CPGE response with disorder via the quasi-particle lifetime or $W^2$ correction, this correction brings different disordered responses to the same level as in Fig. \ref{fig:disorder} $\mathbf{c}$, where $\beta\sim \beta_0 (1-C_WW^2)=\beta_0 (1-\frac{C_{\tau_W}}{\tau_W}), \tau_W\sim \frac{1}{W^2}$ for some coefficient $C_{\tau_W}$.


\section{The CPGE response to rare states in Weyl semimetals} \label{App:D}

\begin{figure}[t!]
\begin{center}
\includegraphics[width = 0.48\textwidth]{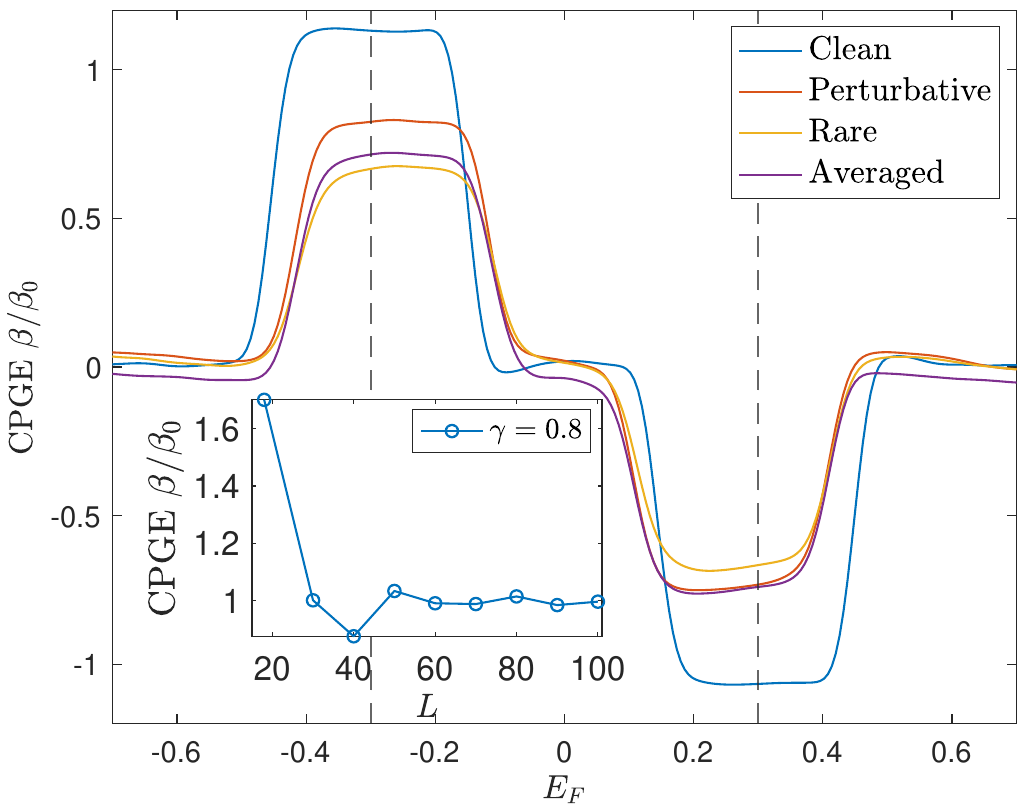}
\caption{The CPGE responses to different disorder realizations at $W=0.5$ with $L=18,N_C=2^9,\omega_1=\gamma=0.3t$. Dashed vertical lines mark the locations $E_F=\pm \omega_1$. The clean, rare and perturbative disorder samples are averaged over $100$ twisted boundary conditions while the averaged result has $100$ disorder realizations. Inset: the CPGE plateau value at various system sizes at $\gamma=0.8$ from Fig. \ref{fig:NCscale}. }
\label{fig:CPGErare}
\end{center}
\end{figure}

It has been shown that rare states in disordered Weyl semimetals give non-perturbative contribution to the DOS and transport \cite{pixley2021rare}, when the two Weyl nodes are at the same energy level $\mu_1=\mu_2=0$ (Fig. \ref{fig:rare}). However, shifting the two Weyl nodes to different energies by introducing finite $\gamma$ in the two-band model makes it hard to identify rare states from squared energy (Fig. \ref{fig:rare} $\mathbf{b}$) as the low energy states are burried in the continuum (Fig. \ref{fig:rare} $\mathbf{a}$) due to the shift. To identify rare states under finite $\gamma$, we first find rare disorder realization from $\gamma=0$ from the squared energy, then slowly breaking the inversion via increasing $\gamma$ at the rare disorder realization. Using the spatial profile of the rare state (eigenstate with power-law decay from it maximal, Fig. \ref{fig:rare} $\mathbf{c}$) and the inverse participation ratio (IPR, Fig. \ref{fig:rare} $\mathbf{d}$)
\begin{equation}
    \mathrm{IPR}(n)=\sum_i |\phi_n(i)|^4,
\end{equation}
we find that the rare state survive up to $\gamma=0.3$. Note that we need the full profile of eigenstates at a finite $\gamma$ and the exact diagonalization is required for the lattice systems, thus we can only use a smaller system size, i.e., $L=18$.

Focusing on the finite $\gamma=0.3$, where the rare state survives, we look at the CPGE response for rare and perturbative disorder realizations in Fig. \ref{fig:CPGErare} and compare them with the clean and averaged results over disorder. 
From Fig. \ref{fig:CPGErare}, there is no qualitative difference between the rare, perturbative, and averaged CPGE response, indicating that the rare states do not have a prominent influence on the CPGE. Note that we are limited to a small system size $L=18$ in order to keep track of rare states using exact diagonalization. Thus, the quantization of the CPGE are less steep than the larger system results in the main text. Also, due to the larger finite size effect, the quantization value is larger in small systems, as shown in the inset of Fig. \ref{fig:CPGErare}.


\bibliography{refs}
 
\end{document}